\pdfoutput=1
\documentclass[10pt,conference]{IEEEtran}
\IEEEoverridecommandlockouts
\usepackage{graphicx}
\usepackage{subcaption}
\usepackage{booktabs}
\usepackage{amsmath,amssymb}
\usepackage{multirow}
\usepackage{xurl}
\usepackage{balance}
\usepackage[numbers,sort&compress]{natbib}
\usepackage{tikz}
\usetikzlibrary{arrows.meta,positioning,fit,backgrounds,calc}
\usepackage{xcolor}
\definecolor{schMandate}{HTML}{6B7280}  
\definecolor{schAdopt}{HTML}{7C3AED}    
\definecolor{schDev}{HTML}{1F3B73}      
\definecolor{schCan}{HTML}{2A8C82}      
\definecolor{schCannot}{HTML}{B45309}   
\newcommand{\Code}[1]{\texttt{#1}}
\makeatletter
\g@addto@macro\normalsize{
  \setlength\abovedisplayskip{4pt plus 2pt minus 1pt}
  \setlength\belowdisplayskip{4pt plus 2pt minus 1pt}
  \setlength\abovedisplayshortskip{1pt plus 2pt}
\setlength\belowdisplayshortskip{3pt plus 2pt minus 1pt}}
\makeatother
\begin{document}
\pagestyle{plain}
\title{AI Writes Faster Than Humans Can Review: A Longitudinal Study of an
  Enterprise ``2$\times$'' Mandate}
\author{\IEEEauthorblockN{
    Hao He,\IEEEauthorrefmark{1}\thanks{Hao He is now at Microsoft; work done while at Carnegie Mellon.}
    Shyam Agarwal,\IEEEauthorrefmark{1}
    Yegor Denisov-Blanch,\IEEEauthorrefmark{2}
    Pavel Azaletskiy,\IEEEauthorrefmark{3}
    Sanmi Koyejo,\IEEEauthorrefmark{2} and
    Bogdan Vasilescu\IEEEauthorrefmark{1}}
  \IEEEauthorblockA{\IEEEauthorrefmark{1}Carnegie Mellon University\quad
    \IEEEauthorrefmark{2}Stanford University\quad
    \IEEEauthorrefmark{3}Independent Researcher}
}
\maketitle
\begin{abstract}
  Enterprises increasingly mandate the use of AI coding tools and set ambitious productivity targets, yet there is little longitudinal evidence on whether such productivity targets have been reached, how, and at what cost.
  In this paper, we present a quantitative case study of a documented enterprise ``2$\times$'' mandate: in June~2025, the CTO of a mid-sized, AI-forward software company set the goal of doubling merged pull requests (PRs) per engineer with AI.
  Using a panel of 802 developers and 196,212 PRs (January~2024--April~2026), we estimate the effect of AI adoption by comparing each developer's PR throughput after they started using AI tools with their throughput before adoption (i.e., a \emph{staggered difference-in-differences} design).
  We find that the company eventually reached the target, but only in the tenth month and not in the way the mandate implied: PR throughput barely changed at the mandate date, increased by about 12\% when each developer started using AI, and then kept increasing as developers accumulated use.
  The target was not met merely by working longer or splitting work: extra working days account for at most one-fifth of the throughput increase, and the added PRs became larger and skewed toward features and bug fixes rather than routine changes.
  The gain is similar across seniority levels and larger in repositories created since 2022. 
  The cost of AI adoption fell on code review, where load per reviewer doubled and automated review exceeded human review in coverage within months of the mandate. 
  We interpret these results as strong evidence that the adoption and sustained usage of AI raises PR throughput in an enterprise setting, with two caveats: throughput is only a proxy for productivity, and the exact size of the effect remains uncertain due to a lack of truly randomized assignment in our setting.
  Our results have implications for organizations under AI transformations and researchers studying the impact of AI on software development productivity.
\end{abstract}
\section{Introduction}
Enterprises are adopting AI coding tools across entire engineering organizations: in 2026, over 70\% of developers reported using AI daily~\cite{stackoverflow2026survey}, and Google's 2025 DORA report put adoption at 90\%~\cite{google2025dora}.
Increasingly, company leaders also mandate adoption.
For example, Shopify made effective AI use ``a fundamental expectation'' of every employee~\cite{cnbc2025shopify}, and Coinbase dismissed engineers who did not adopt the tools~\cite{fortune2025coinbase}.
In this paper, we use \emph{mandate} as a shorthand for this range of top-down commitments to AI-driven productivity, from a stated expectation to enforced adoption, without implying the strictest form in any given case.
These mandates rely on the large payoffs that industry leaders publicly promise, but the credible evidence is far more modest.
Controlled experiments report speedups of 21--56\% on isolated tasks~\cite{DBLP:journals/corr/abs-2302-06590,cui2024effects,DBLP:conf/icse-seip/ParadisGMNMMZFC25}; the largest field study of experienced open-source developers found a 19\% \emph{slowdown}~\cite{becker2025measuring}; a public-sector study did not find a significant change~\cite{DBLP:journals/corr/abs-2509-20353}; and the within-engineer study closest to ours found roughly 40\% more pull requests (PRs) in developers' heaviest-usage weeks~\cite{DBLP:journals/corr/abs-2606-00438}.
Whether a doubling is attainable at all is therefore an open question.
More importantly, these point-in-time estimates provide little evidence on the question an organization faces once it has set a target: what it takes, and how long, to reach it.
Prior qualitative work supplies possible mechanisms, such as the ``productivity pressure paradox'' in which a mandate imposed without enablement undermines its own goal~\cite{DBLP:conf/icse/MillerCUHDSMBB26}, without explicitly tracing the quantitative outcomes across a rollout.
A public target also creates a well-known risk: once a measure becomes a target, it ceases to be a good measure~\cite{strathern1997improving}.
The number an organization reports after a mandate may therefore no longer mean what it meant before.
Answering these questions requires a documented mandate and years of telemetry from an identifiable company, a combination rarely available to researchers.
In this paper, we study one such case: a mid-sized B2B software company with an R\&D organization of several hundred engineers (Section~\ref{sec:methods}, company anonymized per our research agreement).
The company is close to an ideal scenario for an AI mandate: it is AI-forward and pays for commercial AI coding tools with no per-seat or token limits.
If the gains that proponents promise appear anywhere, they should appear here.
In June~2025, its CTO publicly announced a ``2$\times$ mandate'' to double engineering productivity through AI adoption within twelve months, naming merged PRs per engineer per month as the progress metric.
This public commitment gives us a date, a magnitude, a deadline, and a named metric.
Our study window also spans the Opus~4.5 and~4.6 model generations, which were regarded at the time as a step change in agentic coding capability (on vendor benchmarks, in independent evaluations, and in practitioner reviews~\cite{anthropic2025opus45,metr2026timehorizon,willison2025opus45,mowshowitz2025opus45,willison2025yearinllms,anthropic2026opus46}), but whose organizational effects have not been studied longitudinally.
The company granted us research access to its internal AI-tool usage telemetry and PR history, from which we build a panel of 802 developers and 196,212 PRs from January~2024 to April~2026.
Following a longitudinal case study design~\cite{DBLP:journals/ese/RunesonH09}, we estimate the effect of AI adoption by comparing each developer's monthly PR throughput after they start using AI tools with their own throughput before, relative to colleagues whose AI use did not change in the same months.
Developers started using AI at different times, so this is a \emph{staggered difference-in-differences} design~\cite{borusyak2024revisiting,roth2023s}, which we previously used to study the adoption of Cursor in open source~\cite{DBLP:conf/msr/HeMAKV26}.
Specifically, we ask the following research questions:
\begin{itemize}
  \item \textbf{RQ1 (Throughput):}
        How much and how fast did PR throughput rise after the mandate, and what produced the gain?
  \item \textbf{RQ2 (Heterogeneity):}
        Is the gain uniform across developers, the codebase, and model generations?
  \item \textbf{RQ3 (Review and Quality):}
        How did the organization absorb the added throughput, and at what cost to code review and quality?
\end{itemize}
Our findings show that the company reached its target, but slowly and not in the way the announcement implied (RQ1).
Per-capita PR throughput approached $2\times$ only in April~2026, ten months after the announcement, and the change tied to the mandate date itself is small.
Instead, the gain follows each developer's AI adoption: PR throughput rises by about 12\% when a developer starts using AI and then keeps increasing with accumulated use, reaching $1.5\times$ at observed usage under strict calendar-month controls.
Developers did not reach the target merely by working longer or splitting work into more PRs.
The added PRs are larger and contain proportionally more features and bug fixes, and at most a fifth of the gain comes from developers working more days.
To the best of our knowledge, this is among the largest within-developer gains reported from a field deployment of AI coding tools.
However, the rollout was not randomized: developers chose when to adopt and how heavily to use AI.
We therefore interpret the within-developer comparisons, and the way the gain increases with usage, as strong evidence that adopting and continuing to use AI drives the gain, but we cannot pin down the exact size of the causal effect.
The gain is broadly shared but uneven (RQ2): it is statistically indistinguishable across seniority levels, largest in proportional terms for managers, concentrated in newer repositories, barely present in legacy ones, and not clearly larger for AI users after any of the three frontier-model releases.
Finally, the added volume came at a cost to code review (RQ3).
As AI-authored PRs increased, review demand exceeded review capacity.
Each reviewer handled about twice as many PRs, automated review exceeded human review in coverage soon after the mandate, and the human review that remained was increasingly approval without comment.
Revert rates and post-merge fixes stayed flat, so the merged-PR count that defined success recorded the speedup, but not this shift in who reviews the code.
In summary, this paper contributes a longitudinal, instrumented case study of a documented enterprise AI mandate.
It provides within-developer evidence on how long reaching the target took, what produced the gain, who gained, and what it cost in code review.
For organizations that set such targets, our case shows that the gain arrived ten months after the mandate, through each developer's accumulated use and not from the mandate itself, and at the cost of moving code review from humans onto automation.
\section{Related Work}\label{sec:related}
Our study builds on two lines of research: empirical evidence on the effects of AI coding tools on productivity, quality, and code review (\S\ref{sec:rw-evidence}), and the literature on what those effects depend on, in particular developers' accumulated experience with the tools (\S\ref{sec:rw-experience}).
\subsection{Empirical Evidence on AI Coding Tools}\label{sec:rw-evidence}
Most studies of AI coding tools measure velocity or throughput, which is only one dimension of developer productivity~\cite{DBLP:journals/cacm/ForsgrenSMZHB21}, and their results disagree.
Early in-IDE code completion tools showed modest or no measurable effects despite positive developer perceptions~\cite{DBLP:conf/icse/Imai22,DBLP:conf/chi/Vaithilingam0G22}.
In randomized trials, developers then completed tasks 21--56\% faster~\cite{DBLP:journals/corr/abs-2302-06590,cui2024effects,DBLP:conf/icse-seip/ParadisGMNMMZFC25}, and observational studies added evidence of coordination overhead and a shift of effort toward core coding activities~\cite{DBLP:journals/corr/abs-2410-02091,hoffmann2024generative}.
More recent evidence is less optimistic: experienced open-source developers were 19\% \emph{slower} with AI~\cite{becker2025measuring}, and gains went mostly to experienced developers~\cite{daniotti2026who}.
Meanwhile, AI coding tools now include autonomous agents that author entire PRs~\cite{agarwal2026ai,sarkar2025agents,chen2026code}.
Overall, these estimates are point-in-time and of mixed sign, and none of them follows an organization through a rollout.
A second line of work compares throughput gains with their effects on code quality and review, again with mixed results.
AI-generated code shows a persistent velocity--quality tradeoff~\cite{DBLP:conf/msr/HeMAKV26}, introduces technical debt~\cite{liu2026debt}, and weakens code ownership and developer understanding~\cite{martin2026more}, which \citet{storey2026debt} describes as \emph{cognitive} and \emph{intent} debt.
Agent-authored code also churns more~\cite{popescu2026investigating}, and the added output concentrates maintenance on a shrinking pool of experienced reviewers~\cite{xu2025ai}.
In contrast, a controlled experiment did not find any maintainability penalty when later developers evolved AI-co-developed code~\cite{borg2026echoes}.
Code review is both a knowledge-transfer and a defect-detection mechanism~\cite{DBLP:conf/icse/BacchelliB13,DBLP:conf/icse/CzerwonkaGT15}, and it is cognitively demanding~\cite{DBLP:conf/icse/Liang0M24,DBLP:conf/msr/McIntoshKAH14}, so more PRs mean more load on reviewers.
Agent-authored PRs merge faster but are accepted less often and draw more analytic, less social discussion~\cite{njoku2026authors,chung2026collaborator,pinna2026comparing}, and recent work studies why agentic contributions are rejected and how humans and agents split review work~\cite{DBLP:journals/tosem/WatanabeLKRIH26,hindi2026coding,ehsani2026ai,duma2026these,zhong2026human}.
Our study therefore examines where the added work goes after a mandate, especially in code review.
Two studies are closest to ours.
\citet{DBLP:conf/msr/HeMAKV26} use the same staggered difference-in-differences
design to study the adoption of Cursor in open-source projects, and
\citet{DBLP:journals/corr/abs-2606-00438} relate weekly AI usage to output
within each of 16,223 Microsoft engineers, finding roughly 40\% more PRs in
an engineer's heaviest-usage weeks.
Our study differs from these and other field studies in two ways.
First, we follow the same developers for 28 months, from before most of them used AI, through a company-wide mandate, to the point where AI tools author most of their code, whereas prior field studies observe a fixed model over a shorter window.
Second, we date each developer's adoption and intensity of AI use from tool telemetry, so our within-developer comparisons rest on recorded use rather than on self-reports or inferred labels.
\subsection{Complementary Investments and the Productivity
  Paradox}\label{sec:rw-experience}
The wide variation in returns across developers and over time points to a second literature, which indicates that the payoff from a general-purpose technology depends on complementary investments.
For example, the large IT investments of the 1970s to 1990s increased productivity only alongside reorganized workflows, new skills, and decentralized decision making~\cite{brynjolfsson1993productivity,bresnahan2002information}.
Importantly, automation replaces some tasks but increases demand for complementary, judgment-intensive ones~\cite{autor2015why}.
The same pattern appears with AI: gains concentrate among less-experienced workers in writing and customer support~\cite{noy2023experimental,brynjolfsson2023generative}, but performance drops on tasks beyond the model's ``jagged frontier''~\cite{dellacqua2023navigating}.
In coding, the ability to evaluate, debug, and check AI output remains concentrated among experienced developers, and access to AI widens that gap~\cite{fawzy2026prompting}.
Sustained reliance on AI can also erode these skills~\cite{luders2026keeping}.
Qualitative studies find that frequent users treat the tool as a collaborator and invest in learning it: \citet{DBLP:conf/icse/MillerCUHDSMBB26} describe a ``productivity pressure paradox'' in which a mandate that raises expectations without granting time to build skill undermines the gains it seeks.
Because this evidence is mostly qualitative or from other domains, it remains an open question whether developers' gains grow with their accumulated use of AI tools.
Our study tests this quantitatively by separating, within each developer, the one-time jump at adoption from the gain that grows with accumulated use of the tools.
\section{Methods}\label{sec:methods}
We study one organization's AI coding rollout as a quantitative case study~\cite{DBLP:journals/ese/RunesonH09}, following the company across the two years surrounding its ``2$\times$'' mandate.
This section describes the case company (\S\ref{sec:company}), the data we mine (\S\ref{sec:data}), and our analysis approach (\S\ref{sec:approach}).
\subsection{Case Company and the ``2$\times$'' Mandate}\label{sec:company}
The case is a mid-sized commercial B2B software company with an R\&D organization of several hundred engineers.
It offers a software-as-a-service platform that businesses use to support their customers, and its products include AI features.
The codebase combines a long-lived Ruby core, first written in 2012, with newer services in TypeScript, Python, and Go.
Between 2024 and 2026, it grew from about 500 to over 1,000 repositories and from about 4 to 6.5 million lines of code (RQ2 compares repositories created before and after 2022).
The company is AI-forward: it provides commercial AI coding tools without per-seat limits or token budgets, encourages experimentation, and treats AI fluency as a priority.
These conditions make it an unusually favorable site for an AI rollout, so the effects we observe are better interpreted as an upper bound than as a representative industry average.
In June~2025, the CTO sent the R\&D organization an email setting an explicit goal to double engineering productivity through AI adoption over twelve months, naming \emph{merged pull requests per engineer per month} as the metric.
The company granted us research access to its internal, anonymized engineering and AI-tool data for this study.\footnote{The company's role was limited to data sharing; it did not participate in the analysis or writing, and no company employee is an author.}
We use the public ``2$\times$'' commitment only as the documented prompt and calibration point for our analysis; all results below come from the company's internal data.
Developers used two commercial AI coding tools during our study window (Figure~\ref{fig:adoption}): Cursor, with telemetry from February~2025 until its deprecation in February~2026, and Claude Code, with telemetry from July~2025.
In-house agents arrived near the end of our window.
Three frontier models were also released in our window: Sonnet~4.5 with Claude Code~2.0 (September~2025), Opus~4.5 (November~2025), and Opus~4.6 (February~2026), the latter two regarded at release as large advances in agentic coding~\cite{anthropic2025opus45,metr2026timehorizon,willison2025yearinllms,anthropic2026opus46}.
The telemetry does not record which model a developer used, so we do not observe when anyone switched.
Internal uptake comes later than each release and coincides with the company's own rollout, so we treat each release date only as the earliest month in which exposure to the new model could have begun.
\subsection{Data Collection and Processing}\label{sec:data}
\begin{figure}[t]
  \centering
  \includegraphics[width=\columnwidth]{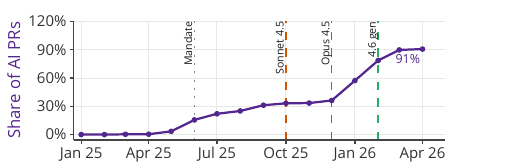}
  \caption{The share of AI-authored PRs climbs from near zero to
    ${\sim}90\%$, with the steepest rise after the Opus~4.5 release.}
  \label{fig:overview-adoption}
\end{figure}

We mine two internal data sources: the company's private GitHub repositories and its AI-tool subscription logs.
\paragraph{Pull-request (PR) and review history}
We extract each PR's title, anonymized author, state, timestamps, size (lines and files changed), labels, commits, comments, and review events.
From these we compute per-PR throughput, review counts, reverts, and two latency measures: \emph{total cycle time}, from first commit to merge or close, which DORA calls \emph{lead time for changes}~\cite{forsgren2018accelerate,google2025dora}, and \emph{review lead}, from the first human review to merge or close.
Bots (AI reviewers and deploy, coverage, and policy bots) account for about $38\%$ of review rows, so we take the first review to be the first non-author \emph{human} review or PR-thread comment.
Suppl.~\ref{app:cycle-decomp} splits total cycle time into coding, pickup, and review phases.
\begin{figure}[t]
  \centering
  \includegraphics[width=\columnwidth]{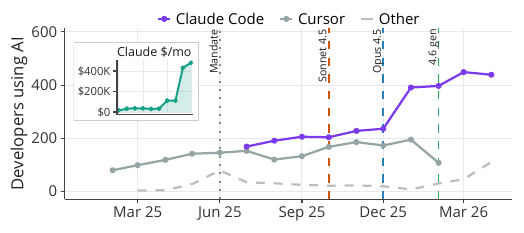}
  \caption{Monthly AI tool users.
    \emph{Inset}: Claude Code token spend.}
  \label{fig:adoption}
\end{figure}
\paragraph{AI-tool usage logs}
From the Cursor and Claude Code logs, we mine per-developer, per-month AI-written and suggested lines, token spend, and acceptance rates, which give us a direct measure of who used AI, when, and how heavily (Figure~\ref{fig:adoption}).
Cursor was the early tool; Claude Code, introduced around the mandate, quickly became dominant.
\paragraph{Identifying AI-authored PRs}\label{sec:pr-ai}
We use the company's own \Code{created-by-ai} PR label, an internal convention for PRs that the company attributes to AI; we do not observe or control how the label is attached.
Throughout the paper, we call PRs with this label \emph{AI-authored}, including in the per-PR regressions (Eq.~\ref{eq:review}).
The label is sparse before the mandate and marks most PRs by the end of our window (Figure~\ref{fig:overview-adoption}).
The label may correlate with other PR characteristics, so we treat comparisons between AI-authored and other PRs as correlational, and reserve causal claims for the developer panel, which relies on the timing of each developer's tool adoption.
The label also captures AI use from tools without usage logs (the \emph{Other} series in Figure~\ref{fig:adoption}).
\begin{figure*}[t]
  \centering
  \includegraphics[width=\textwidth]{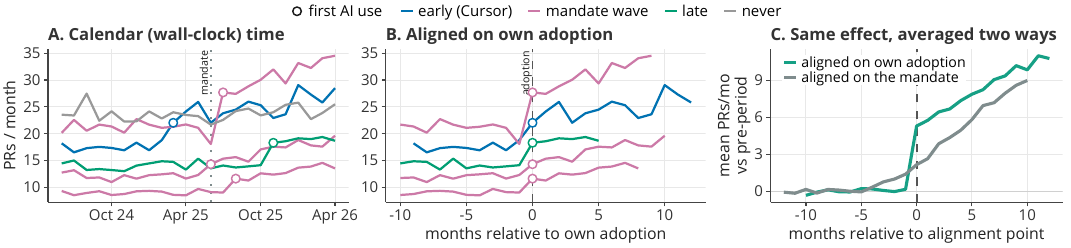}
  \caption{The logic of our design, on simulated data for six developers in
    four adoption cohorts (colors); open circles mark each developer's first AI use.
    \textbf{(A)}~In calendar time, developers adopt AI at staggered dates,
    starting pre-mandate; after the mandate (dotted line), many more developers adopt.
    \textbf{(B)}~Re-centered on each developer's \emph{own} adoption month
    ($\tau=0$), throughput is flat before adoption and clearly jumps and then climbs after.
    That within-developer jump is the effect $\beta_1$ (Eq.~\ref{eq:dev}) estimates.
    It can be interpreted causally under parallel trends---that absent AI, adopters' throughput would have followed the comparison group's (never-adopters)---an assumption supported by the flat pre-adoption trends.
    \textbf{(C)}~The same simulated effect, averaged across a larger population
    for smoothness: the jump is not identifiable in the gradual ramp when the
    curves are aligned on the mandate (a shared date that developers reach at
    different lags relative to their own AI adoption).}
  \label{fig:idlogic}
\end{figure*}
\paragraph{Identifying automated bots}
Following established practice~\cite{dey2020detecting,golzadeh2021ground,he2023automating,lambiase2024motivations}, we flag automated accounts by combining lexical signals (bot and service-account username and email patterns) with behavioral signals (machine-generated branch names, single-repository activity, and templated, low-diversity event bodies), and adjudicate borderline cases manually (Suppl.~\ref{app:bots}).
We also mark a review event as automated when its author is a flagged bot or its body matches a templated-tool pattern.
Review timing therefore starts at the first human review, and per-PR comment counts include only human comments (review-bot comments account for about half of all comment events).
\paragraph{Linking developers, tools, and roles}
We match GitHub users to AI tool logs on company logins.
For most current employees, we also have anonymized HR roles, linked by name and email (${\approx}91\%$ coverage).
For the seniority analysis (Section~\ref{sec:het}), we group role titles into individual contributor, senior, staff, principal, and management tiers.
\paragraph{Resulting datasets}
The final developer-month panel (i.e., one row per developer per month) includes 802 developers from January~2024 to April~2026.
The PR-level dataset contains 196,212 non-bot PRs across 364 active repositories over the same window (Suppl.~\ref{app:data}).
Our estimation sample keeps the $564$ developers with at least three active months ($451$ adopters and $113$ never-adopters), because developers seen for only one or two months contribute no within-developer variation.
\subsection{Analysis Approach}\label{sec:approach}
Our goal is to estimate how much AI changed each developer's PR throughput (i.e., the number of PRs they author per month), and how far that estimate admits a causal interpretation.
The effect may have two components: an immediate jump when a developer first adopts any AI tool, and a further gain as they accumulate use.\footnote{We treat adoption as a single event, a developer's first use of \emph{either} Cursor or Claude Code, because Cursor use was largely pre-mandate and Claude Code post-mandate, leaving the two tools nearly collinear with calendar time.}
We estimate both, because any aggregate change combines them.
Neither comes about at random, however.
Developers who adopted early, or used AI most heavily, differ from the rest: for example, early adopters already had much higher PR throughput before adoption (nearly two standard deviations above other developers).
Simply comparing adopters with non-adopters would therefore confound the effect of the tools with these pre-existing differences, so we never compare the throughput levels of adopters and non-adopters directly.
\paragraph{Staggered difference-in-differences}
Instead, we make each developer their own control, comparing their monthly PR throughput after they start using AI with their throughput in the months before (Figure~\ref{fig:idlogic}A--B).
The $113$ developers who never adopt (Table~\ref{tab:data}) serve a different purpose: they reflect only the shared calendar trend, i.e., the path adopters would have followed had they not started using AI (Figure~\ref{fig:idlogic}B).
Developers adopted at different times, since some took up Cursor well before the mandate while most started using Claude Code in the wave the mandate set off (Figure~\ref{fig:adoption}).
We therefore align each developer on their \emph{own} adoption month rather than on the calendar.
This alignment is known as a \emph{staggered difference-in-differences} design~\cite{roth2023s}.
It is a ``difference in differences'' because the estimate subtracts the change among developers whose AI use did not change in the same months from the change in adopters' throughput around adoption.
It is ``staggered'' because adoption happens at different times for different developers.
The mandate, by contrast, is a single date shared by everyone, so its direct effect cannot be separated from any other organization-wide change at that time.
Aligning developers on the mandate date would also mix developers at different stages of adoption and blur the jump into a gradual ramp (Figure~\ref{fig:idlogic}C).
We therefore treat the mandate as the \emph{catalyst} for adoption and estimate the effect of adopting AI.
We do not estimate a separate direct effect of the mandate.
Our baseline model is the following regression on the developer-month panel:
\begin{equation}\label{eq:dev}
  \log(y_{it}) = \alpha_i + \beta_1\,\text{Adopted}_{it} +
  \beta_2\,\text{Post}_t + \varepsilon_{it},
\end{equation}
where $y_{it}$ is the number of PRs authored by developer $i$ in month
$t$,\footnote{The mandate's metric is \emph{merged} PRs.
  We count \emph{authored} PRs because whether a PR merges also depends on its reviewers.
  Counting only the PRs that eventually merged barely changes the estimate (Section~\ref{sec:throughput}).}
$\alpha_i$ is a \emph{developer fixed effect}, i.e., a separate intercept for each developer that captures all stable differences between developers (so that the estimate uses only each developer's change over time), $\text{Adopted}_{it}$ equals one in months when developer $i$ uses Cursor or Claude Code in the telemetry,\footnote{We base adoption on the usage logs because they are a ground-truth record, rather than on the per-PR \Code{created-by-ai} label, which may miss or misattribute AI assistance.
  Suppl.~\ref{app:prsignal} relaxes this choice and finds the estimate intact, if anything strengthened.}
and $\text{Post}_t$ marks months after the June~2025 mandate.
Specifically, a developer counts as using AI in month $t$ if the telemetry records at least one accepted Cursor line, or any Claude Code lines or spend, in that month; a later month with no recorded use counts as not using AI.
The coefficient $\beta_1$ is thus the average within-developer change in PR throughput on adoption, and $\beta_2$ captures any organization-wide shift after the mandate.
We use this baseline model, with its single $\text{Post}_t$ term, to estimate the mandate's residual effect and to decompose the gain (below).
Our \emph{preferred}, more conservative model replaces $\text{Post}_t$ with \emph{calendar-month fixed effects}, i.e., a separate intercept for each calendar month (Eq.~\ref{eq:monthfe} in Suppl.~\ref{app:throughput}).
These month intercepts capture everything that affects all developers in that month (the mandate, model releases, holidays, and so on), so each adopter is compared only with colleagues observed in the same month (Section~\ref{sec:throughput}).
We also fit a Poisson count model, which, unlike the log model, can keep months with zero PRs (Suppl.~\ref{app:throughput}).
In all models, standard errors are clustered by developer, which allows a developer's monthly observations to be correlated with each other.
The outcome is logged, so coefficients are in \emph{log points}: a coefficient $b$ multiplies the outcome by $e^{b}$ (e.g., $+0.54$ corresponds to $1.72\times$). 
Because $\text{Adopted}_{it}$ is an on/off switch, $\beta_1$ captures a single average step at adoption; whether the gain also grows with use is tested by the usage-tier event study (Figure~\ref{fig:eventstudy}) and by the accumulated-use model described below.
Interpreting $\beta_1$ as a causal effect requires the \emph{parallel trends} assumption: had they not started using AI, developers who adopt in a given month would have seen the same change in throughput as their colleagues, including any change caused by the firm's response to the mandate.
Under this assumption, such shared changes cancel out.
For example, if every developer responded to the mandate by splitting work into more PRs, raising PR counts by 10\% in one month, the estimate for a developer who adopted in that month includes only the increase beyond that 10\%.
The assumption fails if developers were already trending upward before adoption, or if the most eager developers adopted AI \textit{and} changed how they worked (e.g., by splitting PRs or working longer) in the same month.
We consider it plausible here because the CTO set one target for the entire organization, so developers who did not use AI faced the same metric and pressure.
To test the first failure mode, we fit an \emph{event study}, a version of Eq.~\ref{eq:dev} that estimates a separate coefficient for each month before and after adoption: throughput should be flat before adoption and step up afterward (Suppl.\ Figure~\ref{fig:pretrends}).
This check cannot rule out a change in behavior that begins exactly at adoption, so Section~\ref{sec:throughput} additionally checks whether the timing of the gain fits a one-time change in behavior and tests PR splitting and longer hours directly.
We interpret $\beta_1$ as the \emph{total} effect of a developer's adoption, including the changes in that developer's work that follow from it, such as larger PRs that take longer to review (Table~\ref{tab:did}).
We do not control these changes away, because they are part of how adoption pays off, and removing them would subtract part of the effect.
The shift of code review onto automated reviewers is different.
It reached all developers at once, around the mandate.
After June~2025, 82\% of PRs received an automated review, whether or not their author used AI that month.
The calendar-month intercepts therefore absorb it, and it is not part of $\beta_1$ (Section~\ref{sec:process}).
Throughput is also only part of a developer's work.
PR counts correspond to the activity dimension of the SPACE framework~\cite{DBLP:journals/cacm/ForsgrenSMZHB21} and do not include satisfaction, communication, and efficiency.
They also miss work outside the repositories, such as design, meetings, or mentoring.
We use the metric because it is the mandate's own yardstick, and we return to what it omits in Sections~\ref{sec:process} and~\ref{sec:discussion}.
\paragraph{Multiple estimators}
Econometric research shows that this design can mislead when the effect grows over time, because developers who adopted earlier, whose throughput has already increased, serve as controls for those who adopt later~\cite{roth2023s,callaway2021difference}.
We therefore re-estimate the effect with two estimators built for adoption at different times, those of \citet{callaway2021difference} and \citet{borusyak2024revisiting}.
Both use as controls only developers who have not adopted by that month, whether they adopt later or never.
Separately, we fit the baseline model on adopters alone, without the never-adopters.
\paragraph{Distinguishing accumulation from capability}
Next we test whether the gain grows with accumulated AI use, and whether new models added to it.
Every developer could use each new model from the same month, so a model release cannot be separated from anything else that changed in that month, such as the usage ramp or the pressure of the target.
We can, however, test whether developers who used AI gained more after a release than colleagues who did not use AI in the same month.
To do so, we add a developer's total AI-written lines through the previous month, $\text{CumAI}_{i,t-1}$, and one indicator per model release, interacted with $\text{Adopted}_{it}$, to our preferred model:
\begin{equation}\label{eq:cap}
  \begin{aligned}
    \log(y_{it}) = {} & \alpha_i + \gamma_t + \beta_1\,\text{Adopted}_{it} \\
    & {} + \beta\,\log(1{+}\text{CumAI}_{i,t-1}) \\
    & {} + \delta_S\text{Adopted}_{it}\text{Sonnet45}_t + \delta_O\text{Adopted}_{it}\text{Opus45}_t \\
    & {} + \delta_{46}\,\text{Adopted}_{it}\,\text{Gen46}_t + \varepsilon_{it},
  \end{aligned}
\end{equation}
where $\gamma_t$ are the calendar-month intercepts, and each release indicator switches on in the first full month after the release and stays on, so each $\delta$ adds to the previous one.
The month intercepts absorb any change that affected all developers in a month, so each $\delta$ measures only the additional gain of AI users over non-users after that release.
The coefficient $\beta$ is the \emph{gain from accumulated use}: how much throughput increases with accumulated AI use, beyond the adoption jump and the release terms.
We interpret $\beta$ more cautiously than $\beta_1$, because accumulated use is not an exogenous treatment but a choice that varies with workload, task mix, and motivation.
It is also partly mechanical: developers who ship more also accumulate more AI-written lines.
Using the measure from the previous month removes the same-month feedback, and the developer fixed effect absorbs stable productivity differences, but neither rules out that higher throughput causes more use rather than the reverse.
We therefore interpret $\beta$ as an association, i.e., a \emph{dose--response} pattern (the gain rises with the amount of use) that is consistent with a gain compounding from use, and rely on it for the direction of that gain more than for its exact size.
The release terms might also capture a general growth in the gains of AI users over time rather than the effect of the models.
We test this with a \emph{placebo}, a fake release in August~2025, a month without a model launch, where we should not find any effect.
Suppl.~\ref{app:capability} details the release dates and this check.
To split the observed within-developer gain (Section~\ref{sec:throughput}) into the adoption jump and the gain from accumulated use, we add the accumulation term to Eq.~\ref{eq:dev}:
\begin{equation}\label{eq:decomp}
  \begin{aligned}
    \log(y_{it}) = {} & \alpha_i + \beta_1\,\text{Adopted}_{it} +
    \beta\,\log(1{+}\text{CumAI}_{i,t-1}) \\
    & {} + \theta\,\text{Post}_t + \varepsilon_{it}.
  \end{aligned}
\end{equation}
We then evaluate the fitted change at the observed April~2026 usage level and attribute it to a one-time adoption jump ($\beta_1$), a gain from accumulated use ($\beta$), and a residual mandate term ($\theta$).
\paragraph{Downstream models}
For the heterogeneity questions (Section~\ref{sec:het}), we run Eq.~\ref{eq:dev} on subsamples and add interaction terms (adoption $\times$ seniority tier, adoption $\times$ repository cohort), which let the adoption effect differ by group.
The repository-cohort analysis uses a developer-$\times$-cohort-$\times$-month panel, so a developer's adoption is observed separately in older and newer code.
For the per-PR outcomes (Section~\ref{sec:process}), we estimate the
following model:
\begin{equation}\label{eq:review}
  \begin{aligned}
    \log(y_{p}) = {} & \alpha_{i(p)} + \delta_{t(p)} + \beta\,\text{AI}_p +
    \zeta_1 \log(1{+}\text{Size}_p) \\
    & {} + \zeta_2 \log(1{+}\text{Files}_p) + \varepsilon_p,
  \end{aligned}
\end{equation}
where $p$ indexes PRs, $\alpha_{i(p)}$ is an author fixed effect,
$\delta_{t(p)}$ a calendar-month fixed effect, $\text{AI}_p$ the
\Code{created-by-ai} label, and standard errors are clustered by author.
The coefficient $\beta$ is thus the within-author, within-month difference between AI-authored and other PRs, controlling for size because larger changes attract more review attention~\cite{DBLP:conf/icse/SadowskiSCSB18}.
We apply Eq.~\ref{eq:review} to the review and cycle-time outcomes (comment count, review rounds, review coverage, review lead, and total cycle time, with review rounds via Poisson and the rest in logs), and to the binary merge, revert, and any-review outcomes as linear probability models, i.e., linear regressions on a 0/1 outcome whose coefficients are changes in probability.
Trends in review supply across the organization form a single time series, so we report them descriptively rather than model them.
\begin{figure}[t]
  \centering
  \includegraphics[width=\columnwidth]{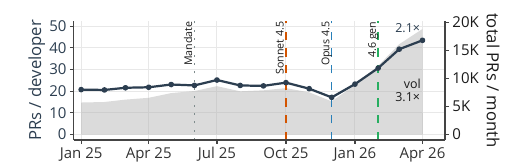}
  \caption{Per-developer throughput (navy, left) grows ${\approx}2\times$
    over the pre-mandate baseline (Jan--Apr~2025), while the company-wide PR
    volume (gray area, right) grows $3.1\times$.}
  \label{fig:overview-throughput}
\end{figure}
\section{Findings}\label{sec:findings}
\subsection{RQ1: A Near-Doubling, Partly Explained by AI Adoption}\label{sec:throughput}
\begin{figure}[t]
  \centering
  \includegraphics[width=\columnwidth]{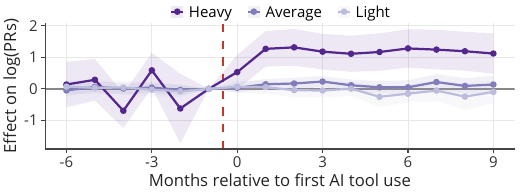}
  \caption{PR throughput rises more for developers who use AI more heavily (a
    descriptive dose--response; Section~\ref{sec:throughput}).
    Event studies on $\log(\text{Monthly PRs})$ with developer and calendar-month fixed effects, estimated separately for each usage tier with never-adopters as the comparison group, relative to the month before a developer's first AI-tool use.
    Bands are 95\% confidence intervals.
  }
  \label{fig:eventstudy}
\end{figure}

Per-capita throughput nearly doubled: mean PRs per active developer rose from 21.2 in the pre-mandate baseline (January--April~2025) to 44.3 in April~2026, a $2.09\times$ increase (Figure~\ref{fig:overview-throughput}).\footnote{Like the other overview figures, this is computed on the estimation sample (the 564 developers with 3+ active months).}$^{,}$\footnote{This counts authored PRs per \emph{active} developer.
  On the mandate's own metric, merged PRs per \emph{engineer} with zero-output months included, the increase is larger ($2.31\times$, Suppl.~\ref{app:throughput}).
  Re-estimating Eq.~\ref{eq:dev} on merged PRs barely changes the adoption coefficient ($+0.354$ to $+0.347$), since the merge rate drifts down only slightly (from 91\% to 85\%).}
The increase does not come from a change in who contributes code, since the 246 developers active both in the baseline and in April~2026 increased their mean throughput by $2.15\times$ (22.9 to 49.3 PRs).
Usage intensity increased even faster.
Per-capita Claude Code token spend among PR authors rose about $20\times$, from \$42 to \$856 per developer-month, between July~2025, when its telemetry starts, and April~2026 (Suppl.\ Figure~\ref{fig:spend}).
Over the same months, PRs per developer rose about $1.7\times$, so tool spend per PR rose more than tenfold, which is a lower bound on the cost of the added throughput because it omits Cursor spend (not dollar-denominated) as well as engineering and review time.
\begin{table}[t]
  \caption{Developer-level difference-in-differences (Eq.~\ref{eq:dev}).
    Developer fixed effects; standard errors clustered by developer.
    Log-point coefficients; approximate percentages via $\exp(\hat\beta)-1$.
    $^{+}p{<}0.10$, $^{*}p{<}0.05$, $^{**}p{<}0.01$, $^{***}p{<}0.001$.}
  \label{tab:did}
  \centering
  \small
  \setlength{\tabcolsep}{2pt}
  \begin{tabular}{@{}lrr@{}}
    \toprule
    Outcome & AI Adopted\phantom{$^{***}$} & Post Mandate\phantom{$^{***}$} \\
    \midrule
    $\log(\text{Monthly PRs})$              & $+0.354^{***}$              &
    $+0.093^{*}\phantom{^{**}}$ \\
    $\log(1{+}\text{Avg PR Size})$          & $+0.317^{***}$              &
    $+0.286^{***}$              \\
    $\log(1{+}\text{Avg Review Time})$      & $+0.111^{*}\phantom{^{**}}$ &
    $-0.029\phantom{^{***}}$    \\
    $\log(1{+}\text{Avg Review Comments})$  & $+0.037^{+}\phantom{^{**}}$ &
    $+0.340^{***}$              \\
    Merge Rate                              & $+0.006\phantom{^{***}}$    &
    $-0.013^{*}\phantom{^{**}}$ \\
    Revert Rate                             & $-0.004^{+}\phantom{^{**}}$ &
    $+0.003\phantom{^{***}}$    \\
    \bottomrule
  \end{tabular}
\end{table}
\begin{table}[t]
  \caption{Decomposing the within-developer throughput gain (Eq.~\ref{eq:decomp}).
    Coefficients on $\log(\text{Monthly PRs})$ with developer fixed effects, standard errors clustered by developer.
    The contribution column gives each term's value at the April~2026 usage level; the total is the implied within-developer change.}
  \label{tab:decomp}
  \centering
  \small
  \setlength{\tabcolsep}{4pt}
  \begin{tabular}{@{}lrr@{}}
    \toprule
    Term & Coefficient & Contribution (log pts) \\
    \midrule
    Adoption jump ($\beta_1$)       & $+0.143^{***}$              & $+0.14$ \\
    Accumulated use ($\beta$)       & $+0.041^{***}$              & $+0.45$ \\
    Residual mandate ($\theta$)     & $-0.05\phantom{^{***}}$     & $-0.05$ \\
    \midrule
    Total (within-developer change) &                             & $+0.54$ \\
    \bottomrule
  \end{tabular}
\end{table}
Within developers, both components of the gain are positive and significant (Table~\ref{tab:decomp}): adopting AI raises a developer's monthly PR throughput, which keeps increasing with accumulated use.
At April~2026 usage, the adoption jump and the gain from accumulated use together predict $+0.59$ log points ($1.81\times$) under the baseline model, or $+0.54$ ($1.72\times$) with the small residual mandate term.
Under the preferred model with calendar-month fixed effects, which capture all organization-wide changes in each month, they predict $+0.38$ ($1.46\times$; adoption jump $+0.11$, accumulated-use slope $+0.025$, each $p<0.05$).
We take this more conservative $1.46\times$ as our main estimate.
A Poisson model that keeps zero-output months confirms the adoption jump ($+0.24$, $p<0.001$; Suppl.~\ref{app:throughput}), so the result is not an artifact of the log transform or of dropping inactive months.
In both models, accumulated use supplies the larger share of the gain and the mandate only a small residual.
As discussed in Section~\ref{sec:approach}, the accumulated-use slope is an association, so we rely on its direction more than its exact size.
Because it is a slope, the implied gain is the value at April~2026 usage and rises with use.
On average, the gain levels off after the first month of use.
In the event study with calendar-month fixed effects, it is $1.32\times$ at three months, $1.27\times$ at six, and $1.41\times$ at nine (Suppl.~\ref{app:pretrends}).
The size of the gain depends instead on how much developers use AI (Figure~\ref{fig:eventstudy}).
Heavy users (top usage quartile) gain about $3\times$ ($+1.1$ to $+1.3$ log points) from month one to month nine after adoption, average users gain 5--26\%, and light users gain nothing.
Before adoption, average and light users do not differ from never-adopters (joint $p=0.52$ and $p=0.68$).
The pre-adoption coefficients of heavy users swing between $-0.7$ and $+0.6$ log points (joint $p<0.001$), but they do not rise toward adoption.
The tiers are defined by \emph{actual} usage after adoption: heavy use may result from high productivity as much as cause it, so we treat the curves for each tier as descriptive.
The raw $2.09\times$ aggregate exceeds all of these within-developer estimates because it also includes the organization-wide calendar trend.
The single pooled adoption estimate (Table~\ref{tab:did}, $+0.354$, about $+42\%$) is smaller than the combined $+0.59$ of the adoption jump and accumulated use, because it summarizes a gain that grows with accumulated use as one step.
It survives calendar-month fixed effects ($+0.201$, $p<0.001$) and dropping never-adopters ($+0.369$), but shrinks under the two estimators that compare each adopter only with not-yet-adopted developers: $+0.132$ ($p<0.10$) under Borusyak imputation and a non-significant $+0.162$ under Callaway--Sant'Anna (Suppl.~\ref{app:throughput}).
These estimators attribute the steep organization-wide ramp of early 2026 to calendar time rather than to AI, which understates an effect that grows with use (Table~\ref{tab:decomp}).
The same bundling of model releases with the calendar means that we cannot separate the part of a release's effect that reached all developers (Section~\ref{sec:het}).
\paragraph{Real work, or a response to the target?}
A public target metric raises the concern that developers reached the target without producing more work.
To test whether the increase in PR throughput reflects real work, we run four checks (Suppl.~\ref{app:goodhart}), all of which support the real work interpretation.
First, a response to pressure would appear soon after the mandate and among all developers; instead, the change tied to the mandate date is small ($+0.09$ log points, Table~\ref{tab:did}), the gain begins only when each developer starts using AI and grows as they accumulate use.
Second, PRs did not get smaller, as splitting would require.
Median PR size nearly doubled from the baseline to April~2026 (35 to 61 lines), and the share of PRs with 10 lines or fewer decreased by nine percentage points.
Lines changed per developer-month also increased faster than PR counts ($+0.81$ versus $+0.57$ log points at April~2026 usage), while files and commits per PR stayed unchanged.
Third, the added PRs are not concentrated in routine work.
Classifying PRs by title, commit prefix, and branch name, we find that feature PRs per developer grow $2.1\times$ and bug-fix PRs $2.6\times$ at April~2026 usage, compared with $1.8\times$ for all PRs, while maintenance PRs (dependency updates, tests, documentation, and cleanup) grow at the overall rate.
Without issue-tracker data, we cannot tell whether the added fixes address older defects or defects in new AI-generated code, but authors do not open fixes after their AI-authored PRs more often than they open features (Section~\ref{sec:process}).
Fourth, longer hours explain only part of the gain.
Developers gain about one active day per month (14.1 to 15.2), and off-hours activity rises from 7\% to 11\%.
However, extra days account for about a third of the adoption jump and only a fifth of the total gain at April~2026 usage.
Because every added PR adds a timestamp, even that share is an upper bound.
None of these checks measures how much the added code accomplishes.
The PR-type classification also relies on titles, which AI tools write in a more uniform format than developers do.
The checks are therefore evidence against metric gaming.
They do not show the value of the added code.
\begin{table}[t]
  \caption{Heterogeneity in the AI adoption effect.}
  \label{tab:tier}
  \centering
  \small
  \begin{tabular}{@{}lrrr@{}}
    \toprule
    Group & $n$ & AI Adopted\phantom{$^{***}$} & approx.\ \% \\
    \midrule
    \multicolumn{4}{@{}l}{\emph{Seniority tier}} \\
    \quad IC          & 163 & $+0.235^{***}$           & $+27\%$ \\
    \quad Senior      & 110 & $+0.352^{***}$           & $+42\%$ \\
    \quad Staff       & 55  & $+0.330^{***}$           & $+39\%$ \\
    \quad Principal   & 16  & $+0.323^{+}\phantom{^{**}}$ & $+38\%$ \\
    \quad Management  & 42  & $+0.619^{***}$           & $+86\%$ \\
    \midrule
    \multicolumn{4}{@{}l}{\emph{Repository cohort}} \\
    \quad Pre-2022 (legacy) & 2{,}763 & $+0.117\phantom{^{***}}$ & $+12\%$ \\
    \quad 2022+ (newer)     & 8{,}121 & $+0.364^{***}$           & $+44\%$ \\
    \bottomrule
  \end{tabular}
\end{table}
\subsection{RQ2: The Gain Is Broadly Shared but Uneven}\label{sec:het}
An average gain can hide differences across people, codebases, and tools, which we examine in turn.
\paragraph{Across seniority levels -- broadly uniform}
Within each engineering tier (Table~\ref{tab:tier}), adoption raises throughput for individual contributors ($+0.24$, $+27\%$), senior engineers ($+0.35$, $+42\%$), and staff engineers ($+0.33$, $+39\%$), each $p<0.001$.
The principal estimate is similar ($+0.32$) but imprecise, with only 16 developers ($p<0.10$).
A model that lets the effect differ by tier does not find significant differences across the IC, senior, staff, and principal tiers, so the gain from AI is fairly uniform across seniority.
Management is the exception, with the largest jump ($+0.62$, ${\approx}+86\%$), consistent with the interpretation that AI is letting managers who had shipped little code contribute more.
\paragraph{Across the codebase -- concentrated in newer repositories}
The picture differs sharply by code age.
On a developer-$\times$-cohort-$\times$-month panel (Table~\ref{tab:tier}, lower block), adoption raises throughput by $+0.36$ ($+44\%$) in repositories created since 2022 but only $+0.12$ ($+12\%$, $p>0.10$) in older ones, and the difference is itself significant ($+0.25$, $p<0.01$).
The newer repositories also use a more modern technology stack, so we cannot fully separate code age from technology.
\begin{figure}[t]
  \centering
  \includegraphics[width=\columnwidth]{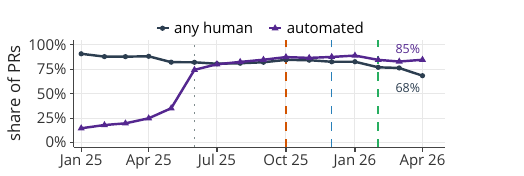}
  \caption{The share of PRs receiving \emph{any} human review (black) falls
    to 68\%, and \emph{automated} review---AI review bots (purple)---exceeds
    it shortly after the mandate, reaching 84\%.}
  \label{fig:overview-review}
\end{figure}
\paragraph{Across model generations -- no added gain for AI users}
Throughput rises with accumulated AI use, but AI users did not gain more than non-users after any model release.
In Eq.~\ref{eq:cap}, the gain from accumulated use is positive and significant ($\beta = +0.024$, $p<0.001$).
None of the three release terms is significant (Sonnet~4.5 $+0.16$, $p=0.06$; Opus~4.5 $-0.01$; Opus~4.6 $-0.25$, $p=0.16$; joint test $p=0.12$), and the August~2025 placebo is of similar size ($+0.09$, $p=0.20$; Suppl.~\ref{app:capability}).
These estimates are imprecise, because few developers worked without AI after each release (36 after the 4.6 generation).
They also cannot capture model effects that reached all developers in the same month, for example through AI review bots or AI tools outside our usage logs (Section~\ref{sec:discussion}).
Per-developer model logs would show who used each new model and when, but they would likely not separate models from accumulated use, because uptake of each release is spread over months and coincides with each developer's rising use of the tools.
\begin{figure}[t]
  \centering
  \includegraphics[width=\columnwidth]{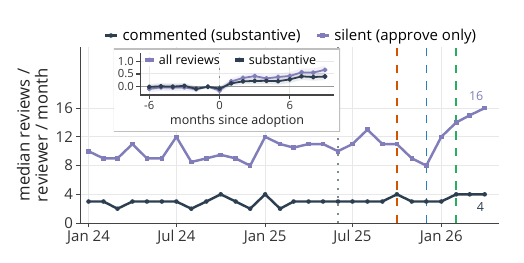}
  \caption{Review activity rises after adoption, faster in light-touch
    (approve-only) than substantive review.
    \emph{Inset:} a within-developer event study around their own AI adoption
    (developer fixed effects, mandate control)
    shows the same gap.}
  \label{fig:reallocation}
\end{figure}
\subsection{RQ3: Adoption Reshaped the Review Process}\label{sec:process}
Reviewers faced the rising PR volume.
As AI-authored PRs grew to ${\sim}90\%$ of the total (Figure~\ref{fig:overview-adoption}), PR volume grew $3.1\times$ over the early-2025 baseline (Figure~\ref{fig:overview-throughput}), while the pool of developers acting as reviewers grew only $1.5\times$, so the load per reviewer roughly doubled ($2.0\times$).
The organization closed this gap partly with more human review but mostly by moving review to automation (Figure~\ref{fig:overview-review}).
The share of PRs receiving at least one human review fell from 89\% to 68\%, while the share receiving an automated AI review increased from ${\sim}19\%$ to ${\sim}84\%$, exceeding human review shortly after the mandate.
Automated review thus replaced part of human review rather than only adding to it.
The remaining human review also thinned.
We distinguish \emph{substantive} reviews, which include a human-written comment, from \emph{silent} approvals, which do not.
The substantive share fell from ${\sim}39\%$ to ${\sim}21\%$ of PRs while silent approvals stayed at ${\sim}50\%$ (Figure~\ref{fig:review-composition}, Suppl.~\ref{app:review-composition}).
Likewise, the median reviewer's commented reviews stayed flat (${\sim}3$ per month) while their silent approvals roughly doubled (Figure~\ref{fig:reallocation}), so most of the added load went to approvals without comment.
\begin{figure}[t]
  \centering
  \includegraphics[width=\columnwidth]{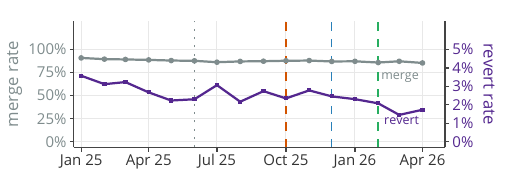}
  \caption{Merge and revert rates stayed stable as review was automated.}
  \label{fig:overview-quality}
\end{figure}
\paragraph{The shift to automation left coarse quality proxies unchanged}
The merge rate stayed essentially flat (Figure~\ref{fig:overview-quality}), so the added volume shipped
rather than remaining open.
At the PR level, controlling for author, month, and change size (Table~\ref{tab:prq}), AI-authored PRs merge at the same rate as other PRs, even though they are \emph{larger} (about 56\% more lines changed post-mandate, $+0.45$ log points; median 66 versus 41 lines).
The merge rate says little about quality, however, since once review becomes lenient or automated, the gate itself has changed.
Signals after merge are more informative, because lenient review that let more defective code through would raise reverts and follow-up fixes.
Instead, the share of PRs that are reverts declined slightly company-wide (Figure~\ref{fig:overview-quality}), and the share of a developer's PRs that are reverts does not increase with adoption (Table~\ref{tab:did}).
Revert PRs per developer do increase, but more slowly than all PRs ($1.4\times$ versus $1.8\times$ at April~2026 usage; Suppl.~\ref{app:prtypes}).
Within 7 or 14 days after an AI-authored PR merges, its author is also no more likely to open a bug fix PR in the same repository than to open a new-feature PR there (differences of 0.3 and 0.6 percentage points, $p=0.78$ and $0.63$; Suppl.~\ref{app:goodhart}).
Without file-level change data this link is coarse, so we report it as suggestive.
These signals warrant care for other reasons as well: the \Code{created-by-ai} label is not randomized and may select toward safer changes, revert shares are diluted as PR volume grows, and both proxies miss defects, incidents, and maintainability (Section~\ref{sec:discussion}).
We interpret these findings as evidence \emph{against} a sudden drop in quality after long-term AI adoption.
\paragraph{Per-PR, the AI penalty is review latency}
If we compare AI-authored and other PRs within author and month, controlling for change size (Eq.~\ref{eq:review}), we find two results (Table~\ref{tab:review}).
First, AI-authored PRs receive about $9\%$ more human comments post-mandate.
Second, post-mandate, AI-authored PRs take about $20\%$ longer from first human review to merge and $22\%$ longer in total cycle time.
These effects are larger and more stable across model variants than the comment effect (Suppl.~\ref{app:cycle-decomp} splits the cycle into phases).
The delay does not appear to reflect rework or evasion: buggier PRs would go through more review rounds, and PRs merged with less scrutiny would draw less human review, yet review rounds are flat and AI-authored PRs are \emph{no less} likely to receive a human review (Table~\ref{tab:review}).
AI-authored PRs instead accumulate more comments without more revision rounds, which points to more reviewer engagement rather than more rework by the author.
\begin{table}[t]
  \caption{Per-PR effects of the \Code{created-by-ai} label (Eq.~\ref{eq:review}).
  Author and month fixed effects, controlling for log size and files; author-clustered SEs.
  Quality and coverage rows are linear probabilities, review rounds Poisson, and review-time rows log-point coefficients ($\exp(\hat\beta)-1$ for percentages).
  Significance levels as in Table~\ref{tab:did}.}
  \label{tab:prq}
  \label{tab:review}
  \centering
  \small
  \begin{tabular}{@{}lrr@{}}
    \toprule
    Outcome & Pooled\phantom{$^{***}$} & Post-mandate\phantom{$^{***}$} \\
    \midrule
    \multicolumn{3}{@{}l}{\emph{Quality (linear probability)}} \\
    P(merged)                         & $+0.006\phantom{^{***}}$    &
    $+0.011^{+}\phantom{^{**}}$ \\
    P(is a revert)                    & $-0.059^{***}$              &
    $-0.067^{***}$              \\
    \addlinespace
    \multicolumn{3}{@{}l}{\emph{Review process}} \\
    $\log(1{+}\text{Comments})$       & $+0.073^{***}$              &
    $+0.083^{***}$              \\
    $\log(1{+}\text{Review Lead})$    & $+0.140^{***}$              &
    $+0.181^{***}$              \\
    $\log(1{+}\text{Total Cycle})$    & $+0.154^{***}$              &
    $+0.202^{***}$              \\
    Review Rounds (Poisson)           & $-0.014\phantom{^{***}}$    &
    $-0.007\phantom{^{***}}$    \\
    Human Review Coverage (LPM)       & $-0.016\phantom{^{***}}$    &
    $-0.002\phantom{^{***}}$    \\
    \bottomrule
  \end{tabular}
\end{table}
This is the bottleneck one would expect when AI speeds up authoring while human review capacity stays fixed: work accumulates downstream.
The \Code{created-by-ai} label may reflect unmeasured complexity that the author, month, and size controls only partly absorb, so these per-PR estimates remain correlational (Section~\ref{sec:pr-ai}).
Moreover, late in the window the company began routing PRs through AI review and auto-approval, which reduced human-review latency for many PRs.
This bypass biases the measured effect toward zero, so our estimate is conservative.
The same bottleneck appears organization-wide.
As PR volume outpaced review capacity, end-to-end cycle time increased and the substantive-review queue never returned to its earlier speed.
The company kept pace by routing work around human review rather than by reviewing faster (Suppl.~\ref{app:org-latency}).
\section{Discussion}\label{sec:discussion}
We estimate a within-developer gain of roughly $1.5\times$ at observed usage under strict calendar-month controls.
This gain is among the largest reported from a field deployment of AI coding tools~\cite{becker2025measuring,DBLP:journals/corr/abs-2606-00438} but it is smaller than the doublings that AI's enterprise proponents claim.
Company-wide throughput did nearly double ($2.09\times$), but the rest of that increase arrived in early 2026 for all developers at once (Section~\ref{sec:throughput}).
We estimate only the effect of each developer's adoption and use.
We do not estimate any effect of AI that reached all developers in the same month, for example better models used by everyone, AI review bots, or AI tools outside our usage logs (Section~\ref{sec:het}).
The effect of AI on the company's throughput is therefore between about $1.5\times$, if none of the shared increase came from AI, and $2.09\times$, if all of it did, and our data cannot place it within this range.
The within-developer gain also grows only as developers accumulate use (Section~\ref{sec:throughput}).
We also measure it in a near-best-case setting (an AI-forward company with uncapped tool budgets, many new repositories, and a strong adoption push), and on a single activity metric that does not capture downstream costs (Section~\ref{sec:process}).
We therefore interpret the result as evidence that a large gain from AI is possible under favorable conditions and with sustained use, rather than that it is typical, immediate, or free.
\subsection{Implications for Practitioners}
\paragraph{Enterprise AI transformation is a continuous process of adaptation}
In our study, throughput barely changed at the mandate date itself ($+0.09$ log points, Table~\ref{tab:did}), the jump at each developer's adoption was about 12\%, and the gain reached about $1.5\times$ only as developers accumulated use (Section~\ref{sec:throughput}).
The gain also concentrated in heavy users and newer repositories, was small and not significant in older repositories (Section~\ref{sec:het}), and came with new ways of working, from managers returning to hands-on coding to a review process rebuilt around AI review bots and auto-approval (Section~\ref{sec:process}).
This pattern matches the IT productivity paradox of the 1970s to 1990s, in which IT paid off only after a lag and alongside reorganized workflows and new skills (Section~\ref{sec:rw-experience})~\cite{brynjolfsson1993productivity,bresnahan2002information}.
We therefore recommend treating a mandate as the start of an organizational transformation.
Organizations should plan over several quarters, invest in intensity of use before demanding the target~\cite{DBLP:conf/icse/MillerCUHDSMBB26}, and pair the target metric with other signals.
A count that becomes a target stops measuring what it was meant to measure (Goodhart's law~\cite{strathern1997improving}).
\paragraph{Human attention and judgment become the bottleneck once AI makes authoring cheap}
In our study, PR volume grew $3.1\times$ while the reviewer pool grew $1.5\times$.
The company closed the gap by routing review to automation, which exceeded human review in coverage within months.
Meanwhile, substantive human review fell from 39\% to 21\% of PRs and silent approvals doubled (Section~\ref{sec:process}).
AI-authored PRs also spent about 20\% longer in human review, and the merged-PR metric did not show any of this.
The automation literature warns that reliance drifts toward misuse when operators cannot keep up with what they supervise~\cite{parasuraman1997humans,lee2004trust}, and developers already describe reviewing AI output as ``an endless stream of AI slop''~\cite{DBLP:journals/corr/abs-2603-27249}.
In this situation, review capacity planning is as important as authoring capacity planning: set targets for review coverage and depth alongside throughput, decide in advance which PRs may be approved without a human review, and monitor which code humans no longer review.
How to allocate scarce human attention across AI output, and how much a human must understand before code ships, remain open questions for both enterprises and researchers.
\subsection{Implications for Research}
Our within-developer estimates add to a literature whose point estimates vary widely, from a 19\% slowdown~\cite{becker2025measuring} to 21--56\% task-level speedups~\cite{DBLP:journals/corr/abs-2302-06590,cui2024effects,DBLP:conf/icse-seip/ParadisGMNMMZFC25} and about 40\% more PRs in heavy-usage weeks~\cite{DBLP:journals/corr/abs-2606-00438}.
Much of this disagreement may reflect differences in time window and setting.
A short window captures only the adoption jump, about 12\% in our data, and misses the later gain from accumulated use.
The gain is also smallest in our older repositories, consistent with the slowdown on mature open-source codebases~\cite{becker2025measuring}, and largest in newer ones, consistent with speedups on self-contained tasks (Section~\ref{sec:het}).
Reconciling them requires longitudinal designs that date exposure from telemetry, and cross-firm or staggered rollouts to separate accumulated use from model capability.
Our code review findings also describe a mechanism through which the quality costs reported elsewhere could accumulate unobserved.
These costs include rising complexity and churn~\cite{DBLP:conf/msr/HeMAKV26,popescu2026investigating,liu2026debt}, diluted ownership~\cite{martin2026more}, and cognitive and intent debt~\cite{storey2026debt}.
Humans examine less of the code, and the automated reviewers that replace them are only beginning to be evaluated~\cite{DBLP:journals/corr/abs-2610-13668,zhong2026human}.
Since DORA and SPACE metrics~\cite{google2025dora,DBLP:journals/cacm/ForsgrenSMZHB21} assume that humans write the code, measuring where human effort goes under agentic development, and what debt accrues without it, is a prerequisite for the next round of productivity studies.
\subsection{Threats to Validity}
\paragraph{Internal validity}
Adoption was not randomized, so a causal interpretation depends on parallel trends, which the flat pre-adoption event study supports but cannot establish (Section~\ref{sec:approach}, Suppl.~\ref{app:pretrends}).
A response to the target that begins exactly at adoption is the main alternative explanation, but never-adopters faced the same target and the checks in Section~\ref{sec:throughput} find little added work time and no PR splitting.
The size of the effect is less certain than its direction ($+0.13$ to $+0.35$ across estimators, Suppl.~\ref{app:throughput}).
The accumulated-use slope is an association, because productive developers accumulate use faster, and model effects are identified only as differences between AI users and non-users, who became rare after each release (Suppl.~\ref{app:capability}).
\paragraph{Construct validity}
PR counts capture only the activity dimension of productivity.
The company applies the \Code{created-by-ai} label, which may favor safer changes, so per-PR contrasts are correlational.
Merge, revert, and follow-up-fix rates are short-horizon proxies that miss defects, incidents, and maintainability, so we do not find any evidence of a sudden drop in quality, but we cannot rule out quality costs.
\paragraph{External validity}
The case is one AI-forward company with uncapped tool budgets, many new repositories, and a public target, so our estimates are better interpreted as an upper bound than as an industry average.
We can estimate factors that were constant for all developers (budgets, training, review policy, how the mandate was communicated) only across companies.
The window also spans rapid change in tools and models, so whether the gain levels off and the review shift persists needs longer panels and replication elsewhere.
\section{Conclusion}
In this paper, we examined how an AI productivity mandate unfolds in practice.
Our longitudinal case study shows that the mandate target is eventually within reach for the case company, but not necessarily in the way implied by the mandate itself.
Instead, the results indicate that organizational AI transformation unfolds through accumulated use and shifts in review practices rather than through the mandate alone.
Human attention and judgment become the new bottleneck of the AI era.
Our study is only one case in a rapidly changing world---we look forward to more research on mandates and AI transformations across diverse organizational contexts, longer time horizons, and beyond AI coding tools.
\textbf{Data Availability.} The data analyzed in this study are proprietary and cannot be shared.
Our statistical analysis code is at \url{https://github.com/CMUSTRUDEL/2x-mandate}.
\textbf{Supplementary Material.}
The supplementary material referenced throughout the paper is in the extended version on arXiv: \url{https://arxiv.org/abs/2607.01904}.
It includes: (1) a summary of the paper's datasets, (2) additional details on automated account detection, (3) robustness checks for code-review-related findings (RQ3), (4) additional checks on the AI adoption definition, the parallel-trends assumption, and alternative estimators (RQ1), (5) robustness checks on whether the observed PR throughput gain reflects real work (testing against alternative interpretations of PR splitting, PR type changes, working time increases, and AI-driven follow-up fixes), and (6) a test of whether AI users gained more than non-users after each model release (RQ2).
\balance
{\small
  \bibliographystyle{IEEEtranN}
  \bibliography{reference}
}
\newpage
\nobalance
\appendices
\begin{center}
  {\large\bfseries Supplementary Materials\\[2pt]
    \normalsize\itshape AI Writes Faster Than Humans Can Review: A Longitudinal
    Study of an Enterprise ``2$\times$'' Mandate}
\end{center}
\vspace{1em}
\section{Dataset Summary}\label{app:data}
Table~\ref{tab:data} summarizes the two datasets used throughout the paper, as described in Section~\ref{sec:data}.
\begin{table}[h]
  \caption{Summary of our datasets (Jan~2024--Apr~2026).}
  \label{tab:data}
  \centering
  \small
  \begin{tabular}{@{}lr@{}}
    \toprule
    Quantity & Value \\
    \midrule
    Developers (panel) & 802 \\
    \quad estimation sample ($\geq$3 active months) & 564 \\
    \quad AI-tool adopters / never-adopters & 451 / 113 \\
    Pull requests (non-bot) & 196{,}212 \\
    \quad AI-authored (\Code{created-by-ai}) & 30.2\% \\
    Repositories (active) & 364 \\
    \bottomrule
  \end{tabular}
\end{table}
\section{Automated-Account Detection}\label{app:bots}
This appendix details the rules we use to flag automated accounts (Section~\ref{sec:data}).
We match each account's username against a list of well-known bots (e.g., \Code{dependabot}, \Code{renovate}, \Code{github-actions}, \Code{copilot}, \Code{vercel}) and against name patterns that signal automation, such as names ending in \Code{[bot]} or \Code{-bot} or resembling CI, auto-merge, or release tooling.
We also inspect the associated email addresses for service-account markers (e.g., \Code{noreply}, \Code{automation}, \Code{notifications}, \Code{service}).
For high-volume accounts that escape these lexical cues, we examine behavioral signals.
On the author side, these are head branches that follow machine-generated templates, activity confined to a single repository, and an absence of authorship labels.
On the reviewer side, they are a high fraction of templated event bodies, low body diversity, and many events with an empty body.
Borderline cases (for instance, humans who invoke a review bot via a mention) are adjudicated manually by inspection.
\section{Review Process Robustness}\label{app:process}
This appendix supports the review findings of Section~\ref{sec:process}.
\subsection{Mean Per-Reviewer Decomposition}\label{app:mean-decomp}
\begin{figure}[h]
  \centering
  \includegraphics[width=\columnwidth]{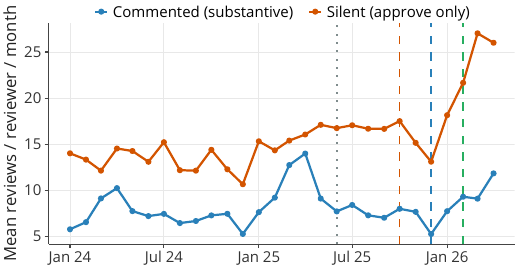}
  \caption{Mean reviews per reviewer, decomposed into commented reviews and
    silent approvals.
    The recurring March--April spikes in the commented series are driven by a few high-volume reviewers and motivate the use of monthly medians in the main text (Figure~\ref{fig:reallocation}) instead.}
  \label{fig:decomp-mean} \end{figure}

Figure~\ref{fig:reallocation} reports the per-reviewer decomposition into commented reviews and silent approvals using monthly \emph{medians}, and Figure~\ref{fig:decomp-mean} shows the same series using \emph{means}.
The means are driven by a small number of very high-volume reviewers and exhibit recurring March--April spikes in the commented series that are not reflected in the median, which is why the main text uses medians.
The qualitative conclusion is unchanged: silent approvals per reviewer rise while commented reviews stay comparatively flat.
\subsection{Human-Review Composition}\label{app:review-composition}
\begin{figure}[h]
  \centering
  \includegraphics[width=\columnwidth]{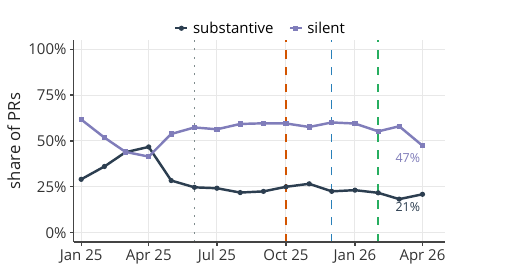}
  \caption{The human-review share decomposed into \emph{substantive} (navy)
    and \emph{silent} approval-only (muted violet) reviews, among non-bot PRs
    on the estimation sample.
    Substantive review decreases while silent approvals stay flat, so the remaining human review is increasingly approval without comment.
    Rules as in Figure~\ref{fig:overview-adoption}.}
  \label{fig:review-composition} \end{figure}

Figure~\ref{fig:overview-review} reports the overall shift from human to automated review, and Figure~\ref{fig:review-composition} decomposes the human share into its \emph{substantive} (with a human-written comment) and \emph{silent} (approval-only) parts.
Substantive review decreases from ${\sim}39\%$ to ${\sim}21\%$ of PRs while silent approvals stay roughly flat (${\sim}50\%$).
The human review that remains is therefore increasingly approval without comment, consistent with the per-reviewer decomposition shown in Figure~\ref{fig:reallocation} of the main text.
\subsection{Per-PR Cycle-Time Decomposition}\label{app:cycle-decomp}
\begin{figure}[h]
  \centering
  \includegraphics[width=\columnwidth]{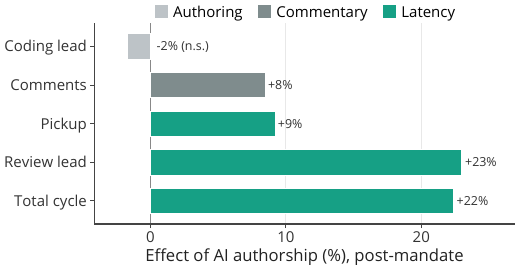}
  \caption{The per-PR effect of AI authorship across the cycle-time phases, controlling
    for size, author, and month (post-mandate).
    AI-authored PRs open slightly sooner after their first commit and draw few extra comments.
    The latency falls in the review-lead phase, while the pickup delay is comparatively small.}
  \label{fig:review} \end{figure}

Total cycle time (Section~\ref{sec:data}) is the sum of three phases: \emph{coding lead} (first commit to PR opened), \emph{pickup} (opened to first human review), and \emph{review lead} (first human review to merge or close).
The boundary between pickup and review lead is the first review or PR-thread comment from a human other than the author, because a bot that responds within seconds would otherwise reduce the measured pickup time to seconds and move the entire wait after opening into review lead.
Table~\ref{tab:phases} reports, for each phase, how much longer AI-authored PRs take under Eq.~\ref{eq:review}, and Figure~\ref{fig:review} visualizes the post-mandate column.
The extra latency of AI-authored PRs concentrates in the review-lead phase; their coding lead is slightly shorter, and the pickup delay is small by comparison with the review-lead phase.
\begin{table}[h]
  \caption{Per-PR effect of AI authorship by cycle-time phase (Eq.~\ref{eq:review});
    log-point coefficients, author and month fixed effects, size controls,
    author-clustered SEs.
    Significance as in Table~\ref{tab:did}.}
  \label{tab:phases}
  \centering
  \small
  \begin{tabular}{@{}lrr@{}}
    \toprule
    Phase & Pooled\phantom{$^{***}$} & Post-mandate\phantom{$^{***}$} \\
    \midrule
    $\log(1{+}\text{Coding Lead})$ & $-0.042^{**}\phantom{^{*}}$ &
    $-0.028^{*}\phantom{^{**}}$ \\
    $\log(1{+}\text{Pickup})$      & $+0.112^{***}$ & $+0.139^{***}$ \\
    $\log(1{+}\text{Review Lead})$ & $+0.140^{***}$ & $+0.181^{***}$ \\
    $\log(1{+}\text{Total Cycle})$ & $+0.154^{***}$ & $+0.202^{***}$ \\
    \bottomrule
  \end{tabular}
\end{table}
\subsection{Organization-Level Latency}\label{app:org-latency}
\begin{figure}[h]
  \centering
  \includegraphics[width=\columnwidth]{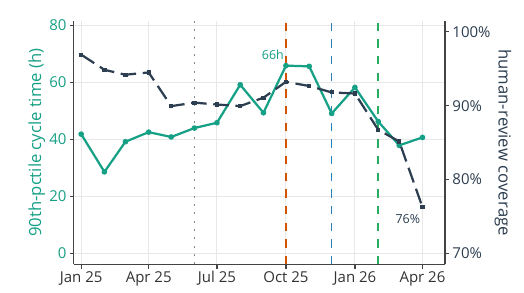}
  \caption{The same bottleneck over time: 90th-percentile end-to-end cycle
    time (teal, left) and human-review coverage (navy, right), both among
    merged PRs.
    Latency increases through 2025 as volume exceeds review capacity, then recedes in 2026 as coverage falls and PRs increasingly bypass human review.
    Monthly, on the estimation sample, excluding bot-authored PRs; vertical rules as in Figure~\ref{fig:overview-adoption}.}
  \label{fig:overview-latency} \end{figure}

The longer per-PR review latency has a counterpart in the organization's aggregate cycle time over the window (Figure~\ref{fig:overview-latency}).
The 90th-percentile end-to-end cycle time increases through 2025 as volume exceeds review capacity, peaking near $66$~hours, and then recedes through early 2026.
However, the recovery comes from a change in which PRs go through review.
When we split PRs by review depth, the 90th-percentile cycle time of PRs that received a \emph{substantive} human review increases to its own peak near $114$~hours and remains around $73$~hours in 2026, above its ${\sim}63$-hour pre-ramp level.
The substantive-review queue itself therefore never returns to its earlier speed.
What falls is the share of PRs routed through it: substantive-review coverage among merged PRs drops from a pre-mandate peak of ${\sim}48\%$ to ${\sim}21\%$, and the fast bypassed lanes pull the aggregate down even as the queue stays slow.
The timing is observational and the monthly tail is noisy, but the pattern corroborates the per-PR result and shows how the case company kept pace: it handled the queue by routing work around substantive human review, rather than by reviewing each PR any faster.
\section{Adoption and Causal Identification}\label{app:causal}
This appendix supports the adoption definition and the parallel-trends assumption underlying the causal interpretation of the adoption jump $\beta_1$ (Section~\ref{sec:approach}).
\subsection{Usage-Intensity Ramp}\label{app:spend}
\begin{figure}[!ht]
  \centering
  \includegraphics[width=\columnwidth]{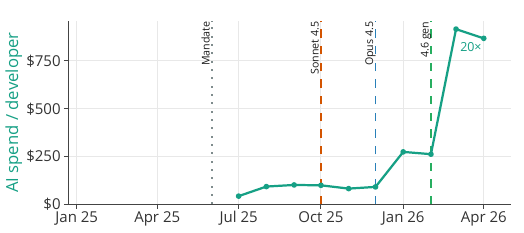}
  \caption{Per-capita Claude Code token spend among active PR authors, in
    dollars per developer-month.
    Spend increases about $20\times$ (\$42 to \$856) over the observation window; the series begins in July~2025 when Claude Code telemetry starts.
    Vertical rules mark the company-wide mandate and the three frontier-model releases in our window.}
  \label{fig:spend} \end{figure}

Figure~\ref{fig:spend} plots the per-capita Claude Code token spend cited in Section~\ref{sec:throughput}.
Spend per developer-month rises roughly $20\times$ between July~2025 and April~2026, from \$42 to \$856, and steepens through the model-release window rather than at any single release date.
Cursor usage is not dollar-denominated, so this is a Claude-only lower bound on the true intensity ramp.
It is also the organization-wide ramp that overlaps with every model release and so prevents release effects from being separated from accumulated use (Section~\ref{sec:het} and Table~\ref{tab:robust}).
\subsection{Adoption Signal Robustness}\label{app:prsignal}
Our primary specification dates a developer's adoption from Cursor and Claude Code usage telemetry, which records tool use directly.
The company's \Code{created-by-ai} label on the PR itself provides an independent adoption signal, and the two signals do not always agree.
Of the 113 developers our telemetry treats as never-adopters, 36 (about a third) authored at least one \Code{created-by-ai} PR.
In addition, 87 of the 451 telemetry adopters had an AI-authored PR before their first telemetry month.
With telemetry alone, we keep these developers in the control group or date their adoption too late, which should, if anything, understate rather than overstate the adoption effect.
To check this, we re-estimate Eq.~\ref{eq:dev} with $\text{Adopted}_{it}$ switched on whenever a developer either has usage telemetry \emph{or} authors a \Code{created-by-ai} PR in that month.
This reclassifies 36 developers from controls to adopters (487 adopters and 77 never-adopters, compared with 451 and 113 under telemetry alone) and moves 87 adoption dates earlier.
The adoption coefficient rises from $+0.354$ to $+0.410$ ($p<0.001$), confirming that the telemetry-only definition is conservative: the developers it misses are genuine adopters whose gains were previously counted in the control baseline.
The residual mandate term falls to near zero, as the adopters identified by the PR label, who started around the mandate, now load on the adoption indicator rather than on the mandate term $\text{Post}_t$.
A separate timing concern is that telemetry is observed only in months when a developer is active in PRs, so tool use in an inactive month dates adoption slightly late.
Assigning adoption dates from the raw tool-use telemetry, ignoring this censoring, changes the coefficient by less than $0.01$, because the recovered months do not have a PR outcome and drop out of the log-count regression anyway.
\subsection{Adoption Event Study (Parallel-Trends Check)}\label{app:pretrends}
\begin{figure}[!ht]
  \centering
  \includegraphics[width=\columnwidth]{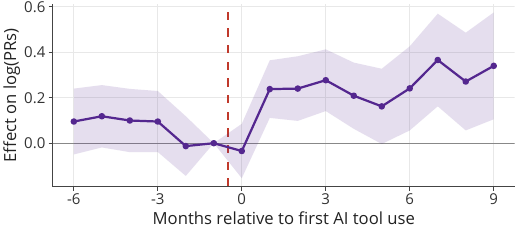}
  \caption{AI-adoption event study on monthly PRs (log scale), with developer and calendar-month fixed effects (Eq.~\ref{eq:es}).
    Coefficients are near zero before adoption, supporting parallel trends, and throughput steps up in the first month after adoption.
    The reference month is $\tau=-1$, and the point at $\tau=-6$ also includes earlier months.
    The shaded band shows the 95\% confidence interval around each monthly estimate.}
  \label{fig:pretrends} \end{figure}

Figure~\ref{fig:pretrends} reports the event study referenced in Section~\ref{sec:approach}.
It replaces the single $\text{Adopted}_{it}$ term of our preferred model (Eq.~\ref{eq:monthfe}) with one indicator per month relative to adoption:
\begin{equation}\label{eq:es}
  \log(y_{it}) = \alpha_i + \gamma_t + \sum_{k=-6,\,k\neq-1}^{12} \delta_k\,\mathbf{1}[t-g_i=k] + \varepsilon_{it},
\end{equation}
where $g_i$ is developer $i$'s first month of AI use, $\gamma_t$ are the calendar-month intercepts, and $\delta_k$ is the change in throughput $k$ months after adoption, relative to the month before.
The month intercepts require never-adopters in the sample, because among adopters alone the month relative to adoption is fully determined by the developer and the calendar month.
We therefore keep the never-adopters as a comparison group and group adopter months before $-6$ or after $+12$ at those endpoints.
Before adoption, the coefficients range from $-0.01$ to $+0.12$, and none is significant (joint test $p=0.28$).
Throughput steps up in the first month after adoption, and the coefficients from one to nine months after adoption range from $+0.16$ to $+0.37$, all significant at $p<0.05$ except month five ($p=0.06$).
This pattern supports the parallel-trends assumption behind the causal interpretation of the adoption jump ($\beta_1$, Section~\ref{sec:approach}).
A simpler version of Eq.~\ref{eq:es}, fit on adopters only with the single $\text{Post}_t$ term in place of $\gamma_t$, gives larger effects after adoption, rising to $1.99\times$ at nine months.
The month intercepts shrink the event-study effects, as they shrink the pooled estimate (Table~\ref{tab:robust}).
The adopters-only version's coefficients before adoption are also small, except for a dip two months before adoption ($-0.18$, $p<0.01$).
Both versions of the event study also use developers who adopted earlier as part of the comparison for those who adopted later.
When the effect of adoption changes over time, such comparisons can bias the estimates~\cite{roth2023s}.
We therefore also estimate the event study with the imputation estimator of \citet{borusyak2024revisiting}, which predicts each adopter's throughput without AI only from months in which developers had not yet adopted or never adopted.
Table~\ref{tab:es-compare} compares the two estimators at selected months after adoption.
The imputation estimator agrees with Eq.~\ref{eq:es} on the shape of the effect.
The coefficients for the months before adoption stay flat (between $-0.02$ and $+0.06$, none significant), throughput steps up in the first month after adoption, and the effect does not rise steadily afterward.
Its estimates are, however, between a third and a half smaller and less precise, and from six months after adoption on they are not significant.
The gap likely reflects the comparison group.
By early 2026, almost all developers had adopted, so the imputation estimator estimates the effect of each calendar month from a small group of not-yet-adopters and never-adopters, whose throughput also rose.
It therefore attributes more of the early-2026 rise to the calendar than Eq.~\ref{eq:es} does, the same pattern as for the pooled estimates (Suppl.~\ref{app:throughput}).
Depending on the estimator, the effect in the months after adoption is thus between about $1.2\times$ and $1.4\times$.
\begin{table}[t]
  \caption{Event-study estimates on $\log(\text{Monthly PRs})$ at selected months after adoption: Eq.~\ref{eq:es} (developer and calendar-month fixed effects) versus the imputation estimator of \citet{borusyak2024revisiting}.
    Significance as in Table~\ref{tab:did}.}
  \label{tab:es-compare}
  \centering
  \small
  \setlength{\tabcolsep}{2pt}
  \begin{tabular}{@{}lrrrr@{}}
    \toprule
    Months after adoption & 1 & 3 & 6 & 9 \\
    \midrule
    Eq.~\ref{eq:es} & $+0.24^{***}$ & $+0.28^{***}$ & $+0.24^{*}\phantom{^{**}}$ & $+0.34^{**}\phantom{^{*}}$ \\
    Imputation & $+0.13^{*}\phantom{^{**}}$ & $+0.20^{**}\phantom{^{*}}$ & $+0.15\phantom{^{***}}$ & $+0.19\phantom{^{***}}$ \\
    \bottomrule
  \end{tabular}
\end{table}
\subsection{Throughput Across Staggered-Adoption Estimators}\label{app:throughput}
The pooled adoption effect of Section~\ref{sec:throughput} is robust to controlling for organization-wide changes over time, and its sensitivity to the choice of estimator is itself informative (Table~\ref{tab:robust}).
Our preferred model replaces the single $\text{Post}_t$ term of Eq.~\ref{eq:dev} with a separate intercept $\gamma_t$ for each calendar month:
\begin{equation}\label{eq:monthfe}
  \log(y_{it}) = \alpha_i + \gamma_t + \beta_1\,\text{Adopted}_{it} + \varepsilon_{it}.
\end{equation}
The month intercepts $\gamma_t$ absorb $\text{Post}_t$ (Section~\ref{sec:approach}).
The preferred version of the decomposition in Section~\ref{sec:throughput} likewise replaces $\theta\,\text{Post}_t$ in Eq.~\ref{eq:decomp} with $\gamma_t$.
The AI-adoption coefficient on $\log(\text{Monthly PRs})$ is $+0.354$ under our baseline model (Eq.~\ref{eq:dev}, with a single $\text{Post}_t$ term) and $+0.201$ ($p<0.001$) once we replace that term with calendar-month fixed effects.
The effect remains highly significant after this reduction of roughly 40\%.
It is $+0.369$ when we drop never-adopters, $+0.132$ under the Borusyak imputation estimator, and shrinks to a non-significant $+0.162$ under Callaway--Sant'Anna.
This pattern localizes the reduction rather than indicating a failure of the result: controlling for calendar time alone leaves the estimate large and highly significant, and the further drop appears only under the estimators that additionally forbid already-adopted developers from serving as controls.
Callaway--Sant'Anna compares each adopter only with developers who have not yet adopted, so the steep organization-wide ramp of early 2026, which lifts adopted and not-yet-adopted developers alike, is attributed to calendar time rather than to AI.
This restriction understates an effect that grows with use (Table~\ref{tab:decomp}), for which a single average step is a poor summary.
The same overlap between the usage ramp and the calendar means that we cannot separate the part of a release's effect that reached all developers (Section~\ref{sec:het}).
\paragraph{Denominator and zero-output months}
The throughput ratio in Section~\ref{sec:throughput} counts authored PRs per \emph{active} developer, whereas the mandate's metric is \emph{merged} PRs per \emph{engineer} per month; conditioning on active developers might in principle inflate the ratio during churn or reactivation.
It does not here.
Recomputing per \emph{engineer}, i.e., counting the zero-output months within each engineer's tenure so that inactive months are part of the denominator, yields a $2.42\times$ increase for authored and $2.31\times$ for merged PRs, at least as large as the per-active ratio, because the pre-mandate baseline contains proportionally more low-output months than the post-ramp endpoint.
The within-developer estimate is likewise robust to including these zero-output months.
On the tenure-spanning panel (11.8\% zero-output months), a Poisson model with calendar-month fixed effects gives an adoption coefficient of $+0.24$ ($p<0.001$).
Conditioning on active developers is thus, if anything, a conservative choice.
\paragraph{Shape of the accumulated-use term}
In the decomposition, we measure accumulated use as $\log(1{+}\text{CumAI}_{i,t-1})$, which assumes that the first AI-written lines add more to throughput than later ones.
This form is a modeling choice, so Table~\ref{tab:shape} refits the preferred decomposition with other forms of the use term.
The implied gain at April~2026 usage ranges from $1.40\times$ for a linear term to $1.59\times$ for a square root, and the square root fits the data slightly better than the logarithm.
We also replace accumulated use with months since adoption.
Both terms based on time since adoption fit less well than any term based on use, and the linear one does not predict throughput ($p=0.36$).
The gain therefore follows how much developers use AI more closely than how long they have used it, while its size at April~2026 usage depends on the form by about $0.1\times$ in either direction.
\begin{table}[!ht]
  \caption{Shape of the accumulated-use term in the preferred decomposition (developer and calendar-month fixed effects).
    Each row replaces $\log(1{+}\text{CumAI}_{i,t-1})$ with another term.
    The gain is the implied effect at April~2026 usage: it adds the adoption jump and the term at its mean April~2026 value among AI users.
    Months counts the months since a developer's first AI use.
    A lower BIC means a better fit.}
  \label{tab:shape}
  \centering
  \small
  \begin{tabular}{@{}lrrr@{}}
    \toprule
    Term & $p$ & Gain & BIC \\
    \midrule
    $\log(1{+}\text{AI lines})$ (main) & $<0.001$ & $1.46\times$ & 23{,}337 \\
    $\sqrt{\text{AI lines}}$ & $<0.001$ & $1.59\times$ & 23{,}318 \\
    AI lines (linear) & $0.002$ & $1.40\times$ & 23{,}339 \\
    $\log(1{+}\text{months})$ & $0.017$ & $1.43\times$ & 23{,}356 \\
    Months (linear) & $0.36$ & $1.33\times$ & 23{,}370 \\
    \bottomrule
  \end{tabular}
\end{table}
\begin{table}[!ht]
  \setlength{\tabcolsep}{2pt}
  \caption{Throughput robustness: AI-adoption coefficient on
    $\log(\text{Monthly PRs})$ across estimators.
    TWFE is the baseline two-way fixed-effects model of Eq.~\ref{eq:dev}; +Month FE replaces its $\text{Post}_t$ term with calendar-month fixed effects (which absorb $\text{Post}_t$); Drop never drops never-adopters; C\&S is the Callaway--Sant'Anna average effect on adopters with not-yet-adopted controls; Borusyak is the imputation estimator with developer and month fixed effects.
    Significance as in Table~\ref{tab:did}.}
  \label{tab:robust}
  \centering
  \small
  \begin{tabular}{@{}lccccc@{}}
    \toprule
    & TWFE & +Mo.\,FE & Drop never & C\&S & Borusyak \\
    \midrule
    $\log(\text{PRs})$ & $+0.354^{***}$ & $+0.201^{***}$ & $+0.369^{***}$ &
    $+0.162$ & $+0.132^{+}$ \\
    \bottomrule
  \end{tabular}
\end{table}
\section{Is the Gain Real Work or a Response to the Target?}\label{app:goodhart}
This appendix reports the checks summarized in Section~\ref{sec:throughput} on whether developers met the ``2$\times$'' target without producing more work.
All analyses use the estimation sample and the within-developer models of Section~\ref{sec:approach}: Eq.~\ref{eq:dev} (``adoption''), its calendar-month fixed-effects variant (``+month FE''), and the decomposition of Eq.~\ref{eq:decomp} evaluated at the April~2026 usage level (``at April~2026 usage'').
Count outcomes use Poisson regressions with developer fixed effects, so coefficients are in log points; standard errors are clustered by developer.
Per-PR size and files are capped at the 99th percentile (winsorized) before monthly sums, because a handful of vendored or generated-file PRs otherwise dominate.
\subsection{Pull-Request Splitting}\label{app:splitting}
If developers met the target by splitting work into smaller PRs, PRs would shrink while lines changed stayed flat.
Table~\ref{tab:splitting} shows the opposite.
Lines changed per developer-month increase faster than PR counts at April~2026 usage ($+0.81$ versus $+0.57$ log points).
The median PR grows from 35 to 61 lines between the pre-mandate baseline (January--April~2025) and April~2026, and the share of PRs with 10 or fewer lines falls from 31\% to 22\%.
Files and commits per PR are unchanged.
The added PRs therefore contain more code, which splitting alone cannot produce.
Lines per PR rise while files per PR do not, which may reflect more verbose AI-generated code; however, we cannot measure how much useful work the added lines actually accomplish.
\begin{table}[!ht]
  \caption{Pull-request splitting: within-developer effects of AI adoption on
    monthly volume and PR shape.
    Volume rows are Poisson (log points); shape rows are OLS on the developer-month average.
    Base and April~2026 columns are means over the estimation sample.
    Significance as in Table~\ref{tab:did}.}
  \label{tab:splitting}
  \centering
  \footnotesize
  \setlength{\tabcolsep}{2pt}
  \begin{tabular}{@{}lrrrrr@{}}
    \toprule
    Outcome & Base & Apr~26 & Adopt. & +Mo.\,FE & Apr~26 use \\
    \midrule
    \multicolumn{6}{@{}l}{\emph{Monthly volume per developer}} \\
    PRs            & 21.2 & 43.4 & $+0.31^{***}$ & $+0.14^{***}$ & $+0.57$ \\
    Lines changed  & 5{,}246 & 12{,}505 & $+0.36^{***}$ & $+0.06$ & $+0.81$ \\
    Files changed  & 120 & 244 & $+0.29^{***}$ & $+0.10$ & $+0.58$ \\
    Commits        & 79 & 150 & $+0.32^{***}$ & $+0.13$ & $+0.41$ \\
    \addlinespace
    \multicolumn{6}{@{}l}{\emph{Shape of a developer's PRs}} \\
    $\log(1{+}\text{lines/PR})$   & 4.78 & 5.28 & $+0.28^{***}$ & $+0.11^{+}$
    & $+0.67$ \\
    $\log(1{+}\text{files/PR})$   & 1.69 & 1.75 & $+0.04$ & $-0.00$ & $+0.10$ \\
    $\log(1{+}\text{commits/PR})$ & 1.40 & 1.40 & $+0.00$ & $+0.01$ & $-0.02$ \\
    Share $\le 10$ lines       & 0.31 & 0.22 & $-0.03^{***}$ & $-0.01$ & $-0.11$ \\
    Share $\ge 500$ lines      & 0.10 & 0.14 & $+0.01$ & $-0.00$ & $+0.05$ \\
    \bottomrule
  \end{tabular}
\end{table}
\subsection{What the Added Pull Requests Contain}\label{app:prtypes}
The company's issue tracker and release records are not part of our data, so we cannot count fixed bugs or shipped features directly.
Instead, we classify each PR by its title, conventional-commit prefix, and branch name, after stripping ticket identifiers, as new functionality, bug fix, maintenance (dependency updates, tests, documentation, cleanup), or revert.
In total, 33\% of PRs do not match any of the patterns and are labeled \emph{other}.
Re-estimating the within-developer models on the monthly count of each type (Table~\ref{tab:prtypes}), we find that all types grow after adoption.
At April~2026 usage, feature PRs per developer grow $2.1\times$ and fix PRs $2.6\times$, compared with $1.8\times$ for all PRs.
Maintenance PRs grow at about the overall rate ($1.9\times$), and the unclassifiable remainder grows least ($1.2\times$).
The added throughput is therefore not concentrated in routine or automatable work such as dependency updates.
Because AI tools write more regular titles than developers do, we also checked whether this shift reflects titling rather than work.
Within author and month, AI-labeled PRs are 10 percentage points more likely to have a classifiable title (78\% versus 63\%), so part of the measured shift could come from more parseable titles.
Among classifiable titles, however, AI-labeled PRs remain 8 percentage points more likely to be fixes post-mandate, so the fix share does not reduce to parseability.
The type mix of PRs \emph{without} the AI label is stable across the window (a fix share of 15\% in both the first half of 2024 and the first half of 2026, and a feature share of 25\% to 27\%), so the company's own conventions did not drift.
Neither check can rule out that AI tools title the same work differently from developers, and no ground truth for what a PR accomplished is available to us for validation.
\begin{table}[!ht]
  \caption{Pull requests by type: within-developer Poisson effects of AI
    adoption on the monthly count of each type.
    Base is the pre-mandate mean count per developer-month, and significance levels are as in Table~\ref{tab:did}.}
  \label{tab:prtypes}
  \centering
  \footnotesize
  \setlength{\tabcolsep}{2pt}
  \begin{tabular}{@{}lrrrr@{}}
    \toprule
    Type & Base & Adopt. & +Mo.\,FE & Apr~26 use \\
    \midrule
    All PRs      & 21.2 & $+0.31^{***}$ & $+0.14^{***}$ & $+0.58$ ($1.8\times$) \\
    Feature      & 5.8  & $+0.36^{***}$ & $+0.17^{**}$  & $+0.76$ ($2.1\times$) \\
    Bug fix      & 2.9  & $+0.48^{***}$ & $+0.22^{**}$  & $+0.96$ ($2.6\times$) \\
    Maintenance  & 3.5  & $+0.31^{***}$ & $+0.14^{*}$   & $+0.62$ ($1.9\times$) \\
    Revert       & 0.7  & $+0.23^{**}$  & $+0.06$       & $+0.36$ ($1.4\times$) \\
    Other        & 8.3  & $+0.19^{***}$ & $+0.10^{*}$   & $+0.21$ ($1.2\times$) \\
    \bottomrule
  \end{tabular}
\end{table}
\subsection{Working Time}\label{app:worktime}
We measure working time from developers' activity timestamps (PRs opened, commits, and reviews written), taking dates and weekends in the company's home time zone and defining off-hours activity as events more than five hours from each developer's own 2024 weekday reference hour.
After adopting AI, developers are somewhat more active (Table~\ref{tab:worktime}).
They have about one more active day per month (14.1 to 15.2, $+0.10$ log points), and off-hours activity rises from 7\% to 11\% of events.
Weekend activity increases slightly in absolute terms (0.3 to 0.6 active weekend days per month), and the daily span from first to last event lengthens by about 40 minutes.
These measures overstate any change in hours, because every additional PR, commit, or review leaves a timestamp.
To see how much of the PR gain extra days explain, we split monthly PRs into active days times PRs per active day and estimate the within-developer effect on each.
The last row of Table~\ref{tab:worktime} fits PRs per active day as a Poisson model of PR counts with $\log(\text{active days})$ as an offset, i.e., holding active days fixed.
Extra days account for about a third of the one-time jump at adoption: of the $+0.31$ log-point jump in PRs, active days contribute $+0.10$ and PRs per active day $+0.23$ (with calendar-month fixed effects the jump is $+0.14$ and splits about evenly).
They explain almost none of the growth after it: at April~2026 usage, active days are up 13\% while PRs per active day are up 59\%, so extra days account for about a fifth of the $+0.57$ total.
Descriptively, between the pre-mandate baseline and April~2026 the average developer's monthly PRs doubled (21 to 43) while active days went from 14 to 15 and PRs per active day from 1.4 to 2.5.
Every added PR adds a timestamp, so the active-days share is an upper bound.
Restricting the events to PRs opened and reviews written, which agents do not author, gives the same picture.
\begin{table}[!ht]
  \caption{Working time: within-developer effects of AI adoption on
    activity-timestamp measures.
    Count rows are Poisson (log points); share and span rows are OLS.
    Base and April~2026 columns are means over the estimation sample.
    Significance levels as in Table~\ref{tab:did}.}
  \label{tab:worktime}
  \centering
  \footnotesize
  \setlength{\tabcolsep}{2pt}
  \begin{tabular}{@{}lrrrrr@{}}
    \toprule
    Outcome & Base & Apr~26 & Adopt. & +Mo.\,FE & Apr~26 use \\
    \midrule
    Active days / month          & 14.1 & 15.2 & $+0.10^{***}$ &
    $+0.08^{***}$ & $+0.13$ \\
    Weekend days / month  & 0.29 & 0.59 & $+0.47^{***}$ & $+0.27^{**}$  & $+0.64$ \\
    Off-hours share    & 0.07 & 0.11 & $+0.02^{***}$ & $+0.02^{***}$ & $+0.03$ \\
    Daily span, hours            & 4.7  & 5.6  & $+0.63^{***}$ &
    $+0.34^{***}$ & $+1.10$ \\
    \addlinespace
    PRs / month                  & 21.2 & 43.4 & $+0.31^{***}$ &
    $+0.14^{***}$ & $+0.57$ \\
    PRs / active day (offset)  & 1.4  & 2.5  & $+0.23^{***}$ & $+0.08^{*}$
    & $+0.46$ \\
    \bottomrule
  \end{tabular}
\end{table}
\subsection{Follow-Up Fixes after AI-Authored Pull Requests}\label{app:followup}
If lenient review let more defective AI-authored code through, the authors of AI-authored PRs should more often have to fix it afterward.
For each merged post-mandate PR, we ask whether the same author opened a bug-fix PR (by the classification above) in the same repository within 7 or 14 days, and compare AI-labeled with other PRs within author and month, controlling for size, in a linear probability model.
Because busier authors open more PRs of every kind, we control for the number of non-fix follow-ups and use follow-up \emph{feature} PRs as a placebo outcome that defect-driven rework should not affect.
AI-labeled PRs are followed by a fix 5.3 percentage points more often within 7 days (base rate 45.5\%) but also by a feature PR 4.8 points more often (base 58.0\%).
The fix-minus-placebo difference is 0.3 percentage points ($p=0.78$) at 7 days and 0.6 points ($p=0.63$) at 14 days.
Authors of AI-authored PRs are thus busier afterward, but not disproportionately busy fixing.
Without file-level change data the link between a PR and a later fix is coarse, so we report this result as suggestive.
\section{Model Releases: AI Users versus Non-Users}\label{app:capability}
This appendix details the release test of Eq.~\ref{eq:cap} (Section~\ref{sec:approach}) and its placebo, and reports the estimates (Table~\ref{tab:ladder}).
\paragraph{Release dates}
Each release indicator switches on in the first full month after the corresponding launch (Section~\ref{sec:company}) and stays on thereafter: October~2025 for Sonnet~4.5, December~2025 for Opus~4.5, and March~2026 for Opus~4.6.
The coefficients $\delta_S, \delta_O, \delta_{46}$ are therefore incremental steps on the previous one.
Entering the release indicators on their own, without the interaction with $\text{Adopted}_{it}$ and without calendar-month intercepts, would compare months before and after each release for all developers at once.
Such a design cannot separate a release from the company-wide usage ramp, and in our data an August~2025 placebo then shows effects as large as the real releases.
The interaction avoids this problem, because the month intercepts absorb the ramp and each $\delta$ compares AI users with non-users in the same month.
\paragraph{Placebo release}
If the gains of AI users grew over time, the release terms would capture that growth.
We therefore replace the three release terms with a single placebo term $\delta_P\,\text{Adopted}_{it}\,\text{Placebo}_t$, where $\text{Placebo}_t = \mathbb{1}[t \ge \text{Aug 2025}]$, a month with no model launch, and leave every other term of Eq.~\ref{eq:cap} unchanged.
If AI users' gains did not grow on their own, $\delta_P$ should be indistinguishable from zero.
\paragraph{Results}
Table~\ref{tab:ladder} reports both models.
None of the release terms is significant at $p<0.05$, and they differ in sign: AI users gained somewhat more than non-users after Sonnet~4.5 ($+0.16$, $p=0.06$), about the same after Opus~4.5, and somewhat less after Opus~4.6 ($-0.25$, $p=0.16$).
The placebo is of similar size to the Sonnet~4.5 term ($+0.09$, $p=0.20$).
The estimates are imprecise because non-users become rare: in the months after Opus~4.6, only 36 developers had a month without recorded AI use (61 developer-months).
The test also cannot detect model effects that reached users and non-users alike, which the month intercepts absorb.
The gain from accumulated use is unaffected by the release terms ($+0.024$ in both models, $p<0.001$).
\begin{table}[t]
  \caption{Model releases: additional gain of AI users over same-month non-users after each release (Eq.~\ref{eq:cap}), with developer and calendar-month fixed effects.
    Each release step is incremental on the previous one.
    We estimate the placebo in a separate regression that replaces the three release terms.
    Significance as in Table~\ref{tab:did}.}
  \label{tab:ladder}
  \centering
  \small
  \setlength{\tabcolsep}{4pt}
  \begin{tabular}{@{}lrr@{}}
    \toprule
    Term & Releases & Placebo \\
    \midrule
    Adopted ($\beta_1$)                         & $+0.080^{+}\phantom{^{**}}$ & $+0.070\phantom{^{***}}$ \\
    Accumulated use ($\beta$)                   & $+0.024^{***}$ & $+0.024^{***}$ \\
    $\times$ Sonnet 4.5 ($\delta_S$)            & $+0.162^{+}\phantom{^{**}}$ & ---\phantom{$^{***}$} \\
    $\times$ Opus 4.5, incr.\ ($\delta_O$)      & $-0.007\phantom{^{***}}$ & ---\phantom{$^{***}$} \\
    $\times$ Opus 4.6, incr.\ ($\delta_{46}$)   & $-0.253\phantom{^{***}}$ & ---\phantom{$^{***}$} \\
    $\times$ Placebo, Aug 2025 ($\delta_P$)     & ---\phantom{$^{***}$} & $+0.093\phantom{^{***}}$ \\
    \bottomrule
  \end{tabular}
\end{table}
\end{document}